\documentclass[aps,prr,a4paper,twocolumn,superscriptaddress,citeautoscript]{revtex4-1}
\usepackage[english]{babel}
\usepackage[T1]{fontenc}
\usepackage{amsmath}
\usepackage{hyperref}
\usepackage{units}
\usepackage[charter]{mathdesign}
\usepackage[pdftex]{graphicx}
\usepackage[dvipsnames]{xcolor}
\usepackage{siunitx}
\usepackage[normalem]{ulem}

\begin{document}
\title{Optical properties of LaNiO$_3$ films tuned from compressive to tensile strain}
\author{I.~Ardizzone}
\affiliation{Department of Quantum Matter Physics, University of Geneva, 24 Quai Ernest-Ansermet, 1211 Geneva 4, Switzerland}
\author{M.~Zingl}
\affiliation{Center for Computational Quantum Physics, Flatiron Institute, 162 5th Avenue, New York, NY 10010, USA}
\author{J.~Teyssier}
\affiliation{Department of Quantum Matter Physics, University of Geneva, 24 Quai Ernest-Ansermet, 1211 Geneva 4, Switzerland}
\author{H.~U.~R.~Strand}
\affiliation{Center for Computational Quantum Physics, Flatiron Institute, 162 5th Avenue, New York, NY 10010, USA}
\affiliation{Department of Physics, Chalmers University of Technology, SE-412 96 Gothenburg, Sweden}
\author{O.~Peil}
\affiliation{Materials Center Leoben Forschung GmbH, Roseggerstrasse 12, A-8700 Leoben, Austria}
\author{J.~Fowlie}
\affiliation{Department of Quantum Matter Physics, University of Geneva, 24 Quai Ernest-Ansermet, 1211 Geneva 4, Switzerland}
\author{A.~B.~Georgescu}
\affiliation{Center for Computational Quantum Physics, Flatiron Institute, 162 5th Avenue, New York, NY 10010, USA}
\author{S.~Catalano}
\altaffiliation[Present address: ]{CIC nanoGUNE, E-20018 Donostia - San Sebastian, Spain}
\affiliation{Department of Quantum Matter Physics, University of Geneva, 24 Quai Ernest-Ansermet, 1211 Geneva 4, Switzerland}
\author{N.~Bachar}
\affiliation{Department of Quantum Matter Physics, University of Geneva, 24 Quai Ernest-Ansermet, 1211 Geneva 4, Switzerland}
\author{A.~B.~Kuzmenko}
\affiliation{Department of Quantum Matter Physics, University of Geneva, 24 Quai Ernest-Ansermet, 1211 Geneva 4, Switzerland}
\author{M.~Gibert}
\altaffiliation[Present address: ]{Physik-Institut, University of Zurich, Winterthurerstrasse190, 8057 Zurich, Switzerland}
\affiliation{Department of Quantum Matter Physics, University of Geneva, 24 Quai Ernest-Ansermet, 1211 Geneva 4, Switzerland}
\author{J.-M.~Triscone}
\affiliation{Department of Quantum Matter Physics, University of Geneva, 24 Quai Ernest-Ansermet, 1211 Geneva 4, Switzerland}
\author{A.~Georges}
\affiliation{Coll\`ege de France, 11 place Marcelin Berthelot, 75005 Paris, France}
\affiliation{Center for Computational Quantum Physics, Flatiron Institute, 162 5th Avenue, New York, NY 10010, USA}
\affiliation{Centre de Physique Th\'eorique, \'Ecole Polytechnique, CNRS, 91128  Palaiseau Cedex, France}
\affiliation{Department of Quantum Matter Physics, University of Geneva, 24 Quai Ernest-Ansermet, 1211 Geneva 4, Switzerland}
\author{D.~van der Marel}\email{dirk.vandermarel@unige.ch}
\affiliation{Department of Quantum Matter Physics, University of Geneva, 24 Quai Ernest-Ansermet, 1211 Geneva 4, Switzerland}
\date{\today}
\begin{abstract}
Materials with strong electronic correlations host remarkable -- and technologically relevant -- phenomena 
such as magnetism, superconductivity and metal-insulator transitions. 
Harnessing and controlling these effects is a major challenge, on which key advances are being made through lattice and strain engineering in thin films and heterostructures, leveraging the complex interplay  between electronic and structural degrees of freedom. 
Here we show that the electronic structure of LaNiO$_3$ can be tuned by means of lattice engineering. 
We use different substrates to induce compressive and tensile biaxial epitaxial strain in LaNiO$_3$ thin films. 
Our measurements reveal systematic changes of the optical spectrum as a function of strain and, notably, 
an increase of the low-frequency free carrier weight as tensile strain is applied. 
Using density functional theory (DFT) calculations, we show that this apparently counter-intuitive effect is due to a change of orientation of the oxygen octahedra.
The calculations also reveal drastic changes of the electronic structure under strain, associated with a Fermi surface Lifshitz transition. We provide an online applet to explore these effects. 
The experimental value of integrated spectral weight below 2 eV is significantly (up to a factor of 3) smaller than the DFT results, indicating a transfer of spectral weight from the infrared to energies above 2 eV. The suppression of the free carrier weight and the transfer of spectral weight to high energies together indicate a correlation-induced band narrowing and free carrier mass enhancement due to electronic correlations.
Our findings provide a promising avenue for the tuning and control of quantum materials employing lattice engineering.
\end{abstract}
\maketitle
\section{Introduction}
Rare-earth nickelates with the chemical formula RNiO$_3$ (with R being a rare earth element)
exhibit a rich structural, electronic and magnetic phase diagram~\cite{torrance1992,garcia1992,catalan2008}. 
In this series of materials La$^{3+}$ has the largest radius, with the effect that LaNiO$_3$ is a paramagnetic metal at all temperatures.  
Intensive research has been motivated by the predicted similarity to cuprates of engineered heterostructures of LaNiO$_3$ and other rare earth nickelates~\cite{chaloupka2008,LNOam2014,Pico,Trilayer,PRMDisa}, efforts which were recently extended to materials that contain elements even further away on the periodic table, such as cobaltates~\cite{Cobaltate}. 
For the smaller rare earths the material is an antiferromagnetic insulator at low temperatures, which passes to an insulating paramagnetic state above the N\'eel temperature and undergoes an insulator to metal transition at a still higher temperature. 
The low temperature insulating state is characterized by two inequivalent nickel sites with electronic states akin to low-spin $4+$ and high-spin $2+$ accompanied by a concomitant rearrangement of the oxygen atoms in a breathing distortion~\cite{mizokawa2000,park2012,mazin2012,johnston2014,subedi2015,seth2017}. 
This "negative $U$" type of bi-stability is understood to arise from an effective attractive intra-atomic electron-electron interaction~\cite{hirsch1985,marel1985,varma1988,strand2014} and  is of interest for materials engineering aimed at superconductivity~\cite{fowlie2018}. 
It is interesting in this context that~\textemdash based on structural and electronic similarities to the cuprates\textemdash~superconductivity has been predicted for strained LaNiO$_3$/LaGaO$_3$ superlattices~\cite{chaloupka2008}. We also note that superconductivity was recently observed in hole doped infinite layer NdNiO$_2$~\cite{li2019}.

Advances in the epitaxial thin film growth on substrates with the perovskite structure have opened an alternative route to rare-earth substitution for controlling the electronic state of transition-metal oxides~\cite{catalano2018}. 
Strain is extensively used in heterostructure engineering of the rare earth nickelates with the composition RNiO$_3$.
The great advantage of this technique is that both compressive and tensile strain conditions can be achieved due to the substrate-imposed epitaxial constraints, producing large structural distortions and  electronic effects. 
Compressive strain mostly modifies the in-plane bond angles while tensile strain is more efficient in modifying the out-of-plane bond angles~\cite{fowlie2019} (see later).
The circumstance that LaNiO$_3$ is exceptional in this class of materials in that it is metallic at all temperatures, motivated us to study the role of strain on the electronic structure of this particular compound.
Unstrained LaNiO$_3$ is a metal with the rhombohedral ($R\overline{3}c$) structure~\cite{zhu2013,guo2018}.
The main effect of strain is to lower the symmetry by reducing the $R\overline{3}c$ space group to a monoclinic $C2/c$ space group (and its supergroups, (see Appendix~\ref{appendix:theory}),
which results in slight orbital polarization, {\em i.e.}, in a splitting between energies of local Ni-centered electronic states.
This degeneracy lifting has a profound impact on the transport and optical properties of thin films.
\begin{figure*}[!ht]
\begin{center}
\includegraphics[width=2\columnwidth]{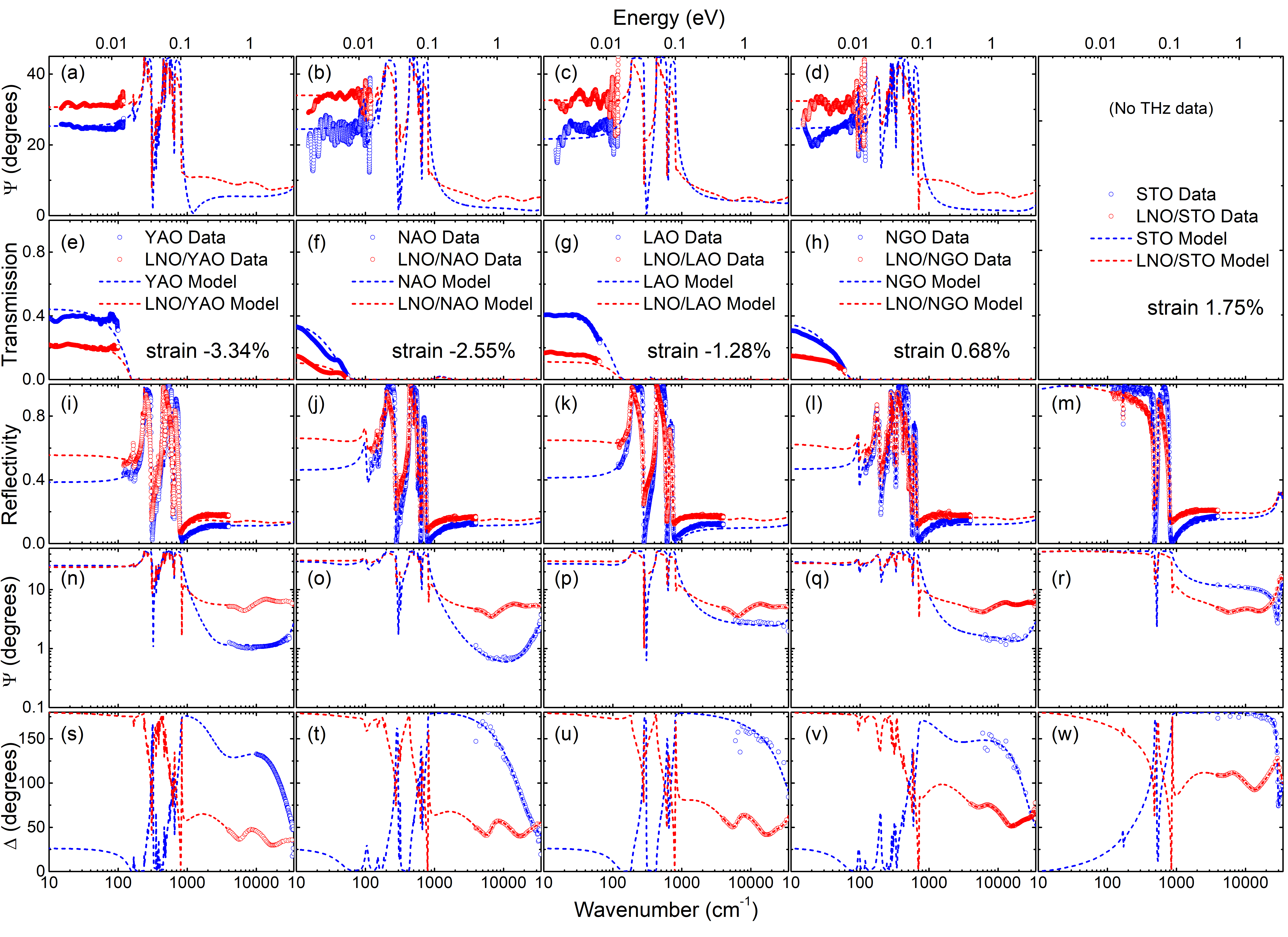}
\caption{\label{fig:optics} 
Full set of optical data at \SI{10}{K} of five LaNiO$_3$ film/substrate combinations at different strains, and simultaneous Drude Lorentz fits to the full data set of each film/substrate combination LaNiO$_3$ on YAlO$_3$(YNO), NdAlO$_3$(NAO), LaAlO$_3$(LAO), NdGaO$_3$(NGO), SrTiO$_3$(STO). Blue dots and curves represent the bare substrate, red dots and curves represent the film/substrate combination. 
{\bf a-d}, $\Psi(\omega)$ of the terahertz ellipsometry measurements. THz ellipsometric data were collected with an angle of incidence of $65^\circ$ ($60^\circ$ for LaNiO$_3$/YAlO$_3$). {\bf e-h}, Transmission in terahertz range.
{\bf i-m}, Infrared reflectivity measurements.
{\bf n-r}, $\Psi(\omega)$ angle of the visible ellipsometry measurements.
{\bf s-w}, $\Delta(\omega)$ angle of the visible ellipsometry measurements. Visible light ellipsometric data were collected with an angle of incidence of $68^\circ$.}
\end{center}
\end{figure*}

Previous optical studies of strained LaNiO$_3$ films have revealed the impact of strain on the optical conductivity and on the enhancement of the electron mass~\cite{ouellette2010,stewart2011}. 
We extend the optical range to energies below 1 meV using a combination of (time-domain and continuous wave) reflection, transmission and ellipsometric spectroscopy. We obtain values of the free carrier mass enhancement that are consistent with the aforementioned reports, and extend those results with a finer set of strains.
Importantly, we relate the observed trends in the strain-dependent optical spectra to Fermi surface (Lifshitz) transitions, and indeed changes in the Fermi surface topology of LaNiO$_3$ due to misfit strain have been previously seen in angle resolved photoemission (ARPES) experiments~\cite{Yoo2015}.
We demonstrate that these transitions have a profound effect on the velocities of Fermi surface states that control the low-frequency conductivity.
\section{Strain dependence of the optical conductivity}
\begin{figure}[t!!]
\begin{center}
\includegraphics[width=\columnwidth]{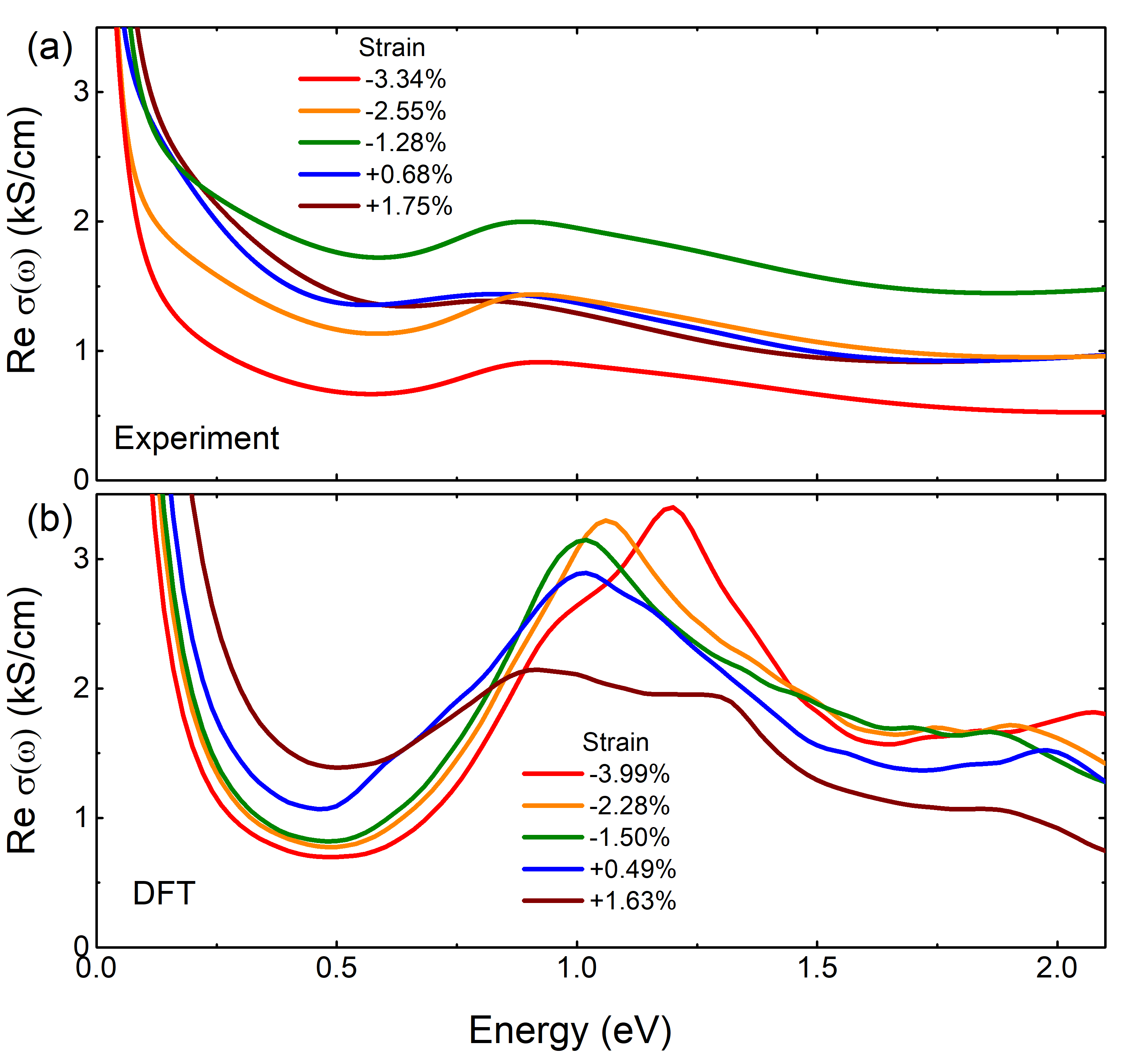}
\caption{\label{fig:sigma}
{\bf a}, Optical conductivity spectra at \SI{10}{K} of LaNiO$_3$ films at different substrate induced in-plane strains indicated in the legend. These spectra roughly fall in two groups: Negative strain (red, orange and green) display a clear peak at 0.8 eV separated from the Drude peak by a minimum at 0.5 eV. Positive strain (blue and brown curves) showing a weak maximum, almost a shoulder, at approx. 0.65 eV. 
{\bf b}, Optical conductivity spectra calculated with density functional theory (DFT) of LaNiO$_3$ films for in-plane strains indicated in the legend and assuming a fixed scattering rate $\hbar/\tau=\SI{60}{meV}$.}
\end{center}
\end{figure}

{We measured the optical spectra of LaNiO$_3$ thin films with substrate-induced strain ranging from highly compressive (-\num{3.34}\%) to moderate tensile (+\num{1.75}\%). Details of the sample preparation are provided in Appendix \ref{appendix:films}. Film specifications are summarized in Table~\ref{table:samples}. 
THz transmission and ellipsometry was measured from \num{5} to \SI{100}{cm^{-1}}; infrared reflectivity from \num{120} to \SI{4000}{cm^{-1}} and near infrared/visible/ultraviolet (NIR/VIS/UV) ellipsometry from \num{4000} to \SI{33000}{cm^{-1}}.
The data were collected at \SI{10}{K} using ultra high vacuum helium flow cryostats.
Due to the high (of order $10^4$) dielectric constant of SrTiO$_3$ no reliable THz ellipsometry and transmission data could be obtained for bare and film-covered SrTiO$_3$.
Ellipsometry provides the ratio $r_p/r_s=\tan(\Psi)e^{i\Delta}$ where $r_p$ and $r_s$ are the reflection coefficient for $p$ and $s$ polarized light. In the THz range we used an angle of incidence of 65$\si{\degree}$ with the surface normal. 
In this range only $\Psi$ could be obtained with sufficient precision. 
In the NIR/VIS/UV range both $\Psi$ and $\Delta$ were measured for angles of incidence with the surface normal varying from 65 to 70$\si{\degree}$ for the different samples (values indicated in Fig.~\ref{fig:optics}). 
For each film/substrate combination the spectra of the bare substrate and of the film-covered substrate were measured. 
We did not observe significant in-plane anisotropy of the optical constants, which confirms the observations in Ref.~\onlinecite{stewart2011}. 
The parameters of a Drude-Lorentz expansion were adjusted with the help of the program RefFit~\cite{kuzmenko2005} to fit the Fresnel equations for transmission, reflection and ellipsometry of stratified media to the full set of data for each film/substrate combination shown in Fig.~\ref{fig:optics}.
These fits were used to generate the complex dielectric function and, associated to this, the optical conductivity.
The corresponding optical conductivities are given in Fig.~\ref{fig:sigma}.}
In order to fit the spectra in the infrared range at low temperature it was necessary to model the optical conductivity at low frequencies as a superposition of a narrow and a broad Drude peak~\cite{nakajima2010,cheng2012}. 
For all temperatures and strains the DC conductivity, $\sigma_1(0)$, obtained on the same samples using four probe measurements fell within the noise level of the optical conductivity at the lowest frequencies of the measured range (i.e. \SI{0.6}{meV}). 
To optimize the result of the Drude-Lorentz fit we therefore used the transport $\sigma_1(0)$ data as an additional constraint. 
The error bars are obtained by varying the relative weights of the different types of spectra (transmission, reflection, ellipsometry) in the fit procedure.

\begin{figure}[t!!]
\begin{center}
\includegraphics[width=\columnwidth]{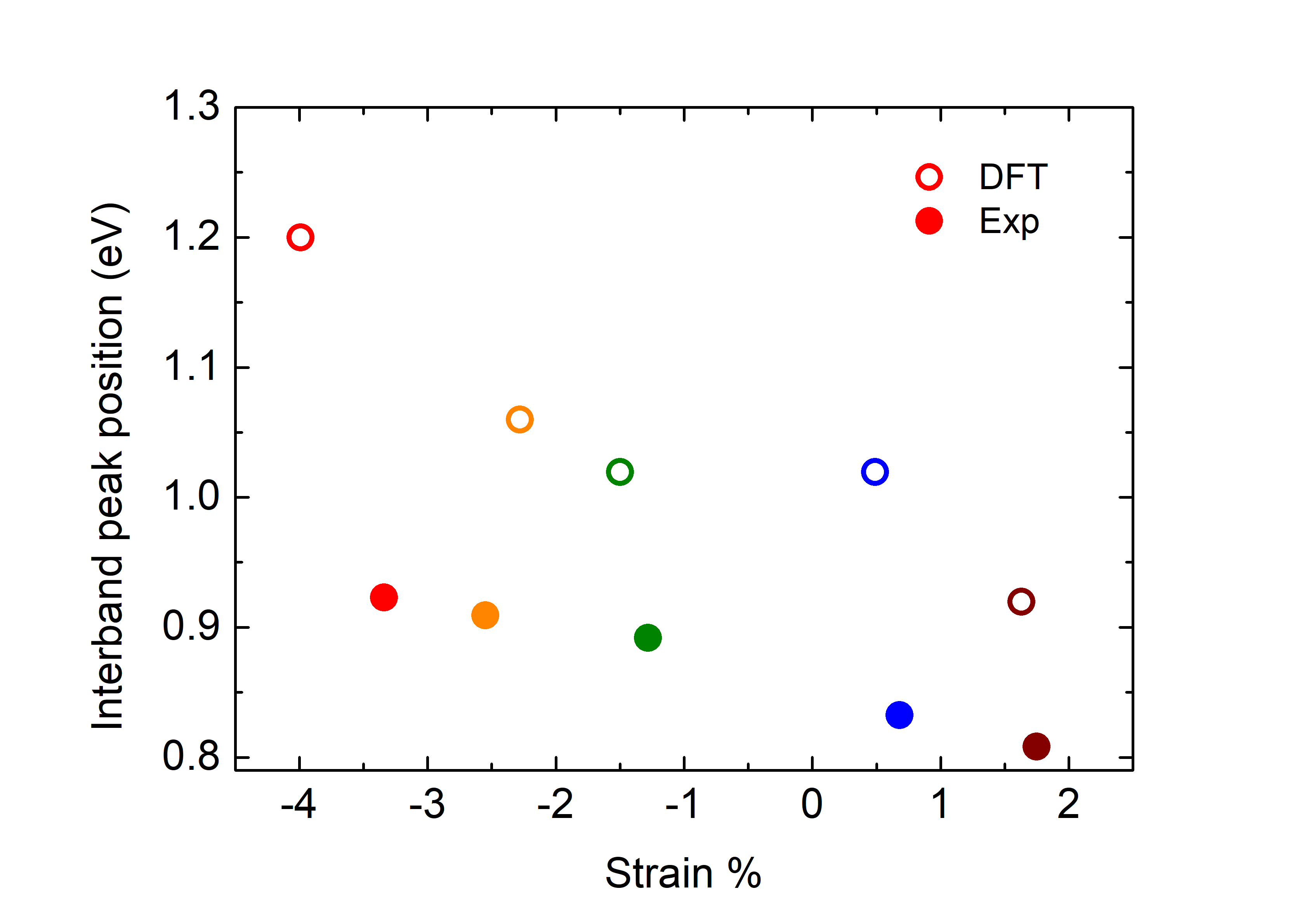}
\caption{\label{fig:peakmax}
Position of the maximum of the peak of the optical conductivity at~$\sim 1$~eV, in experiments and in DFT calculations. 
An overall decrease of the peak position is observed as strain is varied  from the most compressive to the most tensile. 
}
\end{center}
\end{figure}

The optical conductivities at \SI{10}{K} are shown in Fig.~\ref{fig:sigma}{\bf a}.
We observe a broad peak around \SI{1}{eV}, which has been reported previously for LaNiO$_3$~\cite{ouellette2010,stewart2011} 
and for the metallic high temperature phase of RNiO$_3$ with R = Sm, Nd~\cite{ruppen2015}. 
The spectrum at low frequencies is dominated by the zero energy mode due to the free carrier response. 
For frequencies above approximately \SI{0.1}{eV} the optical conductivity falls well above the tail of the Drude peak. 
Empirically this line shape resembles a superposition of a broad Drude peak ($\hbar/\tau$ = \SI{0.3}{eV}) and a narrow one ($\hbar/\tau <$ \SI{0.1}{eV}). 
Given that the two-Drude fit only captures the gross features in the infrared part of the spectrum, and finer details are buried by the strong phonon peaks of substrate and film, it is not possible to provide a detailed analysis of the scattering rate. 
We note however that the superposition of a broad and a narrow Drude peak is expected for a Fermi-liquid~\cite{berthod2013} 
for which the quasi-particle scattering rate has the following frequency dependence~\cite{gotze1972,berthod2013}: 
$1/\tau=\alpha +\beta \omega^2$. 
Indeed, the narrow peak corresponds to the frequency range dominated by the constant part of the scattering rate, while the broader one is due to the frequency-dependent term.
This observation supports the well-established Fermi-like $T^2$ temperature dependence of the resistivity of bulk and thick film LaNiO$_3$ below \SI{50}{K}~\cite{Sreedhar1992,son2010}. In thin films the exponent is usually lower \cite{Scherwitzl2011,scherwitzl2012}. Recently LaNiO$_3$ films on LSAT substrates were found to show $T$-linear resistivity below 1 K as a result of magnetic impurity-induced short-range antiferromagnetic fluctuations~\cite{Liu2020}. We note that the peak position, indicated in Fig.~\ref{fig:peakmax}, decreases gradually going from the most compressive to the most tensile strain.
\par 

\begin{figure}[t!!]
\begin{center}
\includegraphics[width=\columnwidth]{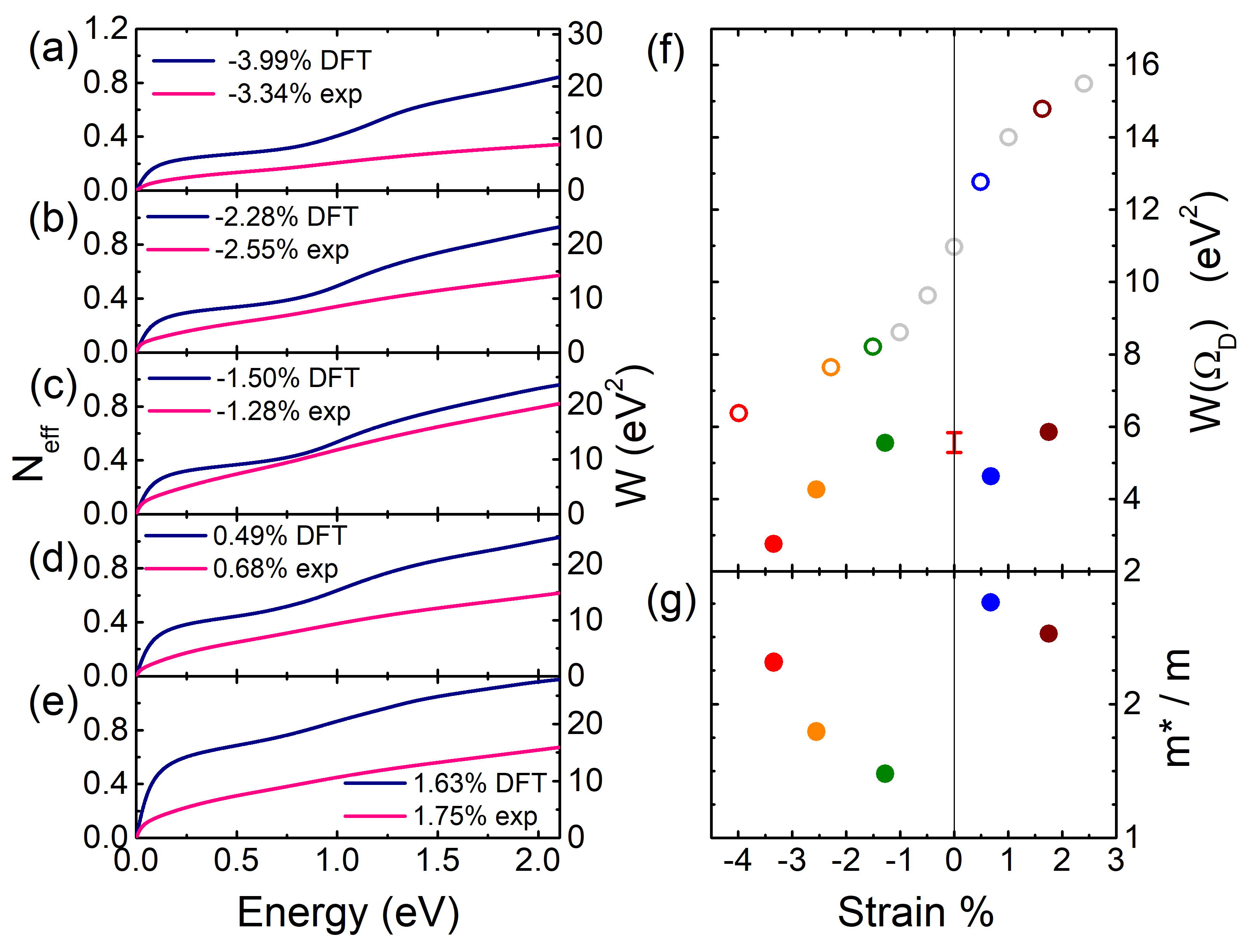}
\caption{\label{fig:neff}
{ 
{\bf a-e} Optical spectral weight $W$ as a function of frequency from experiment at \SI{10}{K} and DFT calculations for each strain value.
{\bf f}, Experimental (closed symbols), and DFT (open symbols) evolution of the free carrier spectral weight  $W(\Omega_D)$ of LaNiO$_3$ films as a function of strain. 
For the theoretical values of $-3.99\%$ strain the calculated crystal structure is $I4/mcm$; 
for all other strain values the structure converged to $C2/c$. 
The error bar applies to all experimental points and was determined by repeating the Kramers-Kronig analysis after multiplying the reflectivity with $1\pm 0.05$.
{\bf g}, Strain dependence of the effective mass according to Eq.~\ref{eq:Wstar}. 
}}
\end{center}
\end{figure}

We also determined the integrated spectral weight, defined as:
\begin{equation}
W(\Omega)  =  \frac{4\pi \hbar^2e^2}{m_e V_{fu}} N_{eff}(\Omega) = 8\hbar^2 \int_{0}^{\Omega} \mbox{Re}~\sigma(\omega) d\omega 
\label{eq:W}
\end{equation}
where $V_{fu}$ is the volume of one formula unit, $\hbar$ the reduced Planck constant, $e$ and $m_e$ the electron charge and mass respectively. The resulting experimental and theoretical effective number of carriers $N_{eff}(\Omega) $ are displayed in Fig.~\ref{fig:neff}{\bf a}-{\bf e}. 
By virtue of the $f$-sum rule the limiting value of $N_{eff}(\Omega)$ for $\Omega\rightarrow\infty$ corresponds to the number of electrons  per formula unit, and the spectral weight in this limit corresponds to the squared plasma frequency of all (core and conduction) electrons: $W(\infty)=\hbar^2\omega_p^2$.  
Of particular interest is the spectral weight of the narrow Drude peak, corresponding to $W(\Omega_D)$,  taking $\Omega_D=\SI{0.35}{eV}$ which is well above the Drude width $1/\tau^*$ but small enough not to include interband transitions and mid-infrared spectral weight. The values of  $W(\Omega_D)$ as a function of strain are displayed in Fig.~\ref{fig:neff}{\bf f}. A key observation is that the experimental data display an overall increase of the Drude weight by approximately a factor of two when going from the most compressive to the most tensile strain. It increases from most compressive to -1.3$\%$ strain, suddenly drops between -1.3$\%$ and +0.68$\%$ and then increases again when the tensile strain increases to 1.75\%.
\section{Strain dependence of the electronic structure}

We have performed electronic band structure calculations using density functional theory (DFT), and display the computed optical conductivities of this compound in the aforementioned range of strain values in Fig.~\ref{fig:sigma}{\bf b} (for calculation details see Appendix~\ref{appendix:theory}). 
We see that the overall frequency dependence of the experimental data in Fig.~\ref{fig:sigma}{\bf a}
as a function of strain is qualitatively well accounted for by the DFT results.

\par

\begin{figure}[t!!]
\begin{center}
\includegraphics[width=\columnwidth]{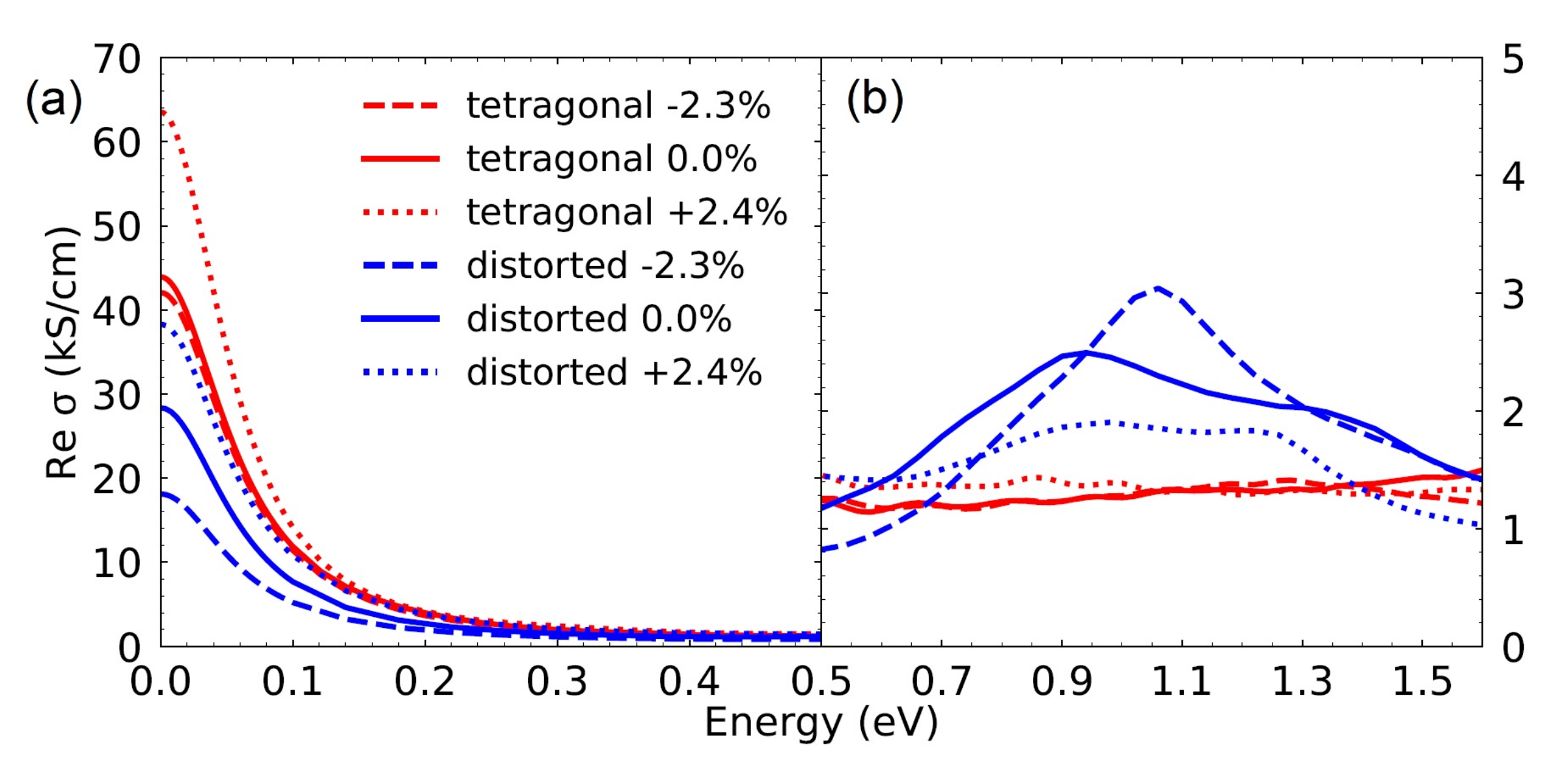}
\caption{\label{fig:DFT_sigma} Calculated DFT optical conductivity {\bf  a}, below and {\bf  b}, above 0.5 eV of tetragonal structures without oxygen-octahedra rotations and tilts (red) compared to the fully distorted structures (blue) at -2.3\% (dashed), 0.0\% (solid) and +2.4\% (dotted) strains assuming a fixed scattering rate of $\hbar / \tau = \SI{60}{meV}$.}
\end{center}
\end{figure}

By comparing to calculations for a (hypothetical) $P4/mmm$ tetragonal structure without NiO$_{6}$ octahedral distortions (see Fig.~\ref{fig:DFT_sigma}), we obtain direct evidence that the broad peak around \SI{1}{eV} 
is due to optical inter-band transitions, being allowed as a result of the band-backfolding occurring in the distorted structure.

The lowering of symmetry and the presence of two Ni atoms per unit cell allow for optical transitions that would otherwise be forbidden in the undistorted tetragonal structure. 
The decrease of the position of this peak from compressive to tensile strain displayed in Fig.~\ref{fig:peakmax} is also qualitatively captured by the DFT calculations.

We compare in Fig.~\ref{fig:neff} the computed (DFT) values of the spectral weight as a function of strain with the experimental data. A key observation is that experiment and computation both display an overall increase of the Drude weight $W(\Omega_D)$ (Fig.~\ref{fig:neff}{\bf f}) by approximately a factor of two when going from the most compressive to the most tensile strain.  
This trend may appear as counter-intuitive: one might have expected an overall decrease of the inter-atomic in-plane hopping strength, and hence of $W(\Omega_D)$, when atoms get farther away from one another under tensile strain. 
Our DFT calculations correspondingly yield the opposite trend of a decrease of the out-of-plane Drude weight and conductivity from compressive to tensile (see Appendix~\ref{appendix:theory} and Fig.~\ref{fig:tau_strain}), a prediction which we leave for future experimental confirmation.
Explaining these apparently counter-intuitive trends is one of the main emphases of our work.

\begin{figure}[t!!]
\begin{center}
\includegraphics[width=\columnwidth]{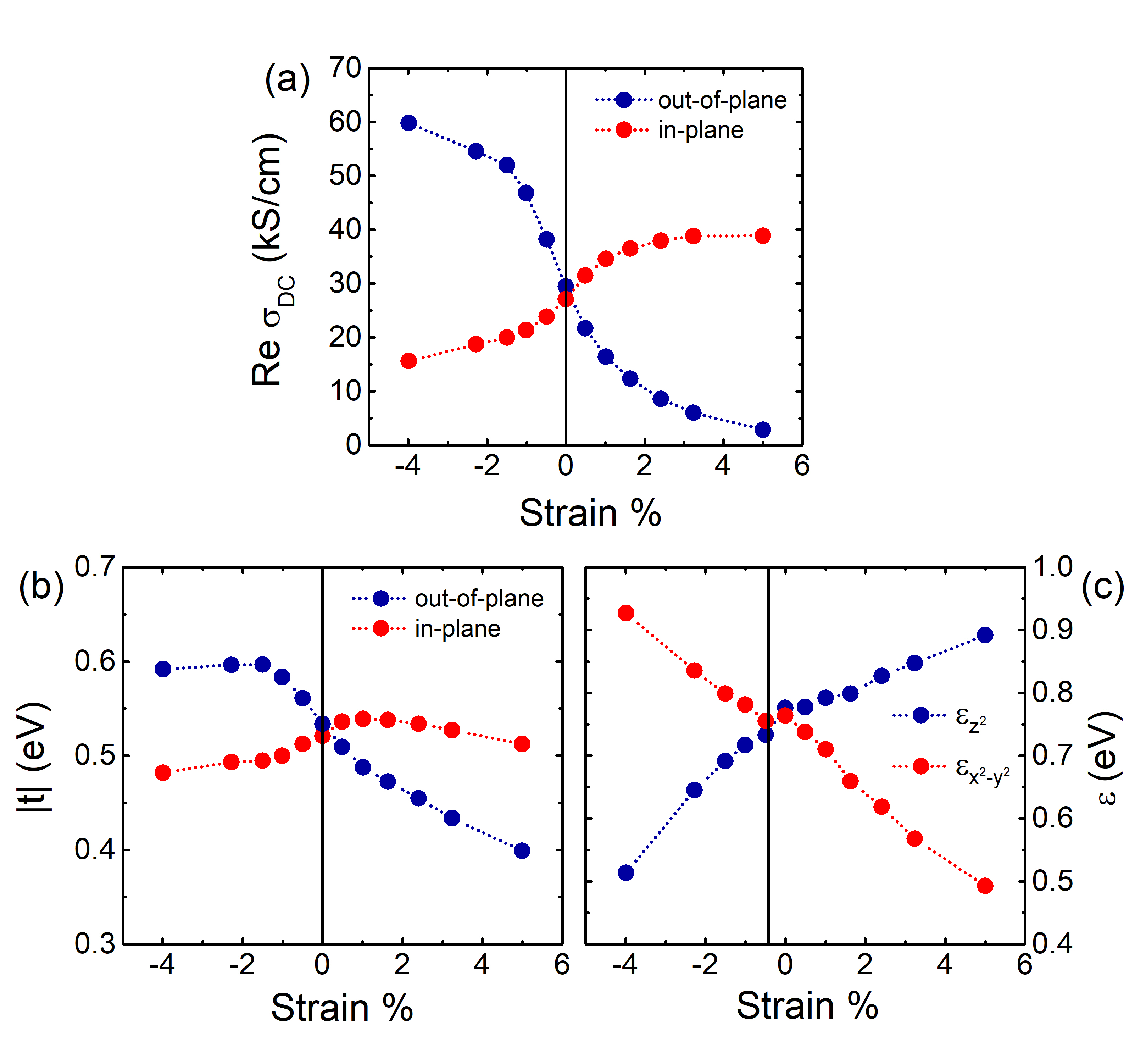}
\caption{\label{fig:tau_strain} 
{\bf a}, Calculated strain dependence of the in-plane (red) and out-of-plane (blue) DC conductivity using a fixed scattering rate of $\hbar / \tau = \SI{60}{meV}$. 
{\bf b}, Wannier model effective in-plane $t_{ab}$ (red) and out-of-plane $t_c$ (blue) hopping amplitudes, and 
{\bf c} on-site energies of the $d_{z^2}$ (blue) and $d_{x^2-y^2}$ (red) Wannier orbitals.}
\end{center}
\end{figure}
A striking feature of the DFT calculations  -- absent from the experimental data -- is the sharp increase between -1\% and +1\% strain. The experimental values even show a slight dip at  +0.7\% strain. 
\footnote{Since the measured optical response depends on the film thickness, one may suspect that this discrepancy could result from assuming the wrong value for the film thickness, or from the presence of a dead ({\em i.e.} insulating) layer. However, the thicknesses of these films are accurately known from X-ray diffraction. Since even for the insulating phase of RNiO$_3$ the spectral weight in the visible part of the spectrum is not so different from the metallic phase~\cite{ruppen2015}, the assumption of a dead layer cannot explain the low value of N$_{eff}$(2eV) for the tensile samples. Moreover, transport experiments on films on LaAlO$_3$ as a function of film thickness have demonstrated that the films switch as a whole from metal to insulator at a critical thickness~\cite{fowlie2017}.
}
The other observation from Fig.~\ref{fig:neff}{\bf f} is that the experimental Drude weight is systematically smaller than the DFT computed value, leading to effective masses of around two (see Fig.~\ref{fig:neff}{\bf g} and further discussion below): this is a hallmark of correlation effects. 

\par 

These two observations are not limited to the low-frequency Drude weight but apply to the integrated spectral weight $W(\Omega)$ for all measured frequencies, as shown in Fig.~\ref{fig:neff}{\bf a-e}.
The experimental optical conductivity at negative strain shows a clear peak at $\sim 0.8$ eV separated from the Drude peak by a minimum at $\sim 0.5$ eV. For positive strain there is a weak maximum, almost a shoulder, at $\sim 0.65$ eV.  In the theoretical spectra we also see that  the minimum at $\sim 0.5$ eV is constant for negative strain, and starts to fill up for positive strain. 
While experiment and DFT agree on this qualitative aspect, the spectral weight evolves differently in theory and experiment when passing from negative to positive strain. 

\par 
To analyze this further it is of interest to make an estimate of the mass enhancement by comparing the measured and calculated spectral weights of the Drude peak: 
\begin{equation}
\frac{m^*}{m} = 
\frac{W_{\text{DFT}}(\Omega_D)}{W_{\text{EXP}}(\Omega_D)} \ .
\label{eq:Wstar}
\end{equation}
The mass enhancement, shown in Fig.~\ref{fig:neff}{\bf g}, is between about $1.5$ and $3.0$, consistent with previous reports~\cite{ouellette2010,stewart2011,king2014}. Based on dynamical mean-field theory calculations, this is the expected range of mass enhancements resulting from the on-site Hubbard and Hund's rule interactions in these compounds~\cite{ouellette2010,stewart2011,deng_prb_2012,peil2014,nowadnick2015,Yoo2015}. 
Most members of the RNiO$_3$ family exhibit a metal-insulator transition which has been characterized as a `site selective' Mott transition~\cite{park2012,subedi2015}. While LaNiO$_3$ remains metallic at all temperatures, it is overwhelmingly natural to assume that the Hubbard $U$ and Hund's $J$-interaction play an important role in the physical properties of this material. One of the most typical consequences of the on-site interaction is a transfer of optical spectral weight from the free carrier response of the Ni $3d$ band to high energy~\cite{basov2011}. Consequently most of the missing free carrier spectral weight is recovered in the oxygen-$2p$ to Ni-$3d$ transitions which span over approximately 10~eV. The fact that in Fig.~\ref{fig:neff}{\bf a}-{\bf e} the spectral weight difference between experiment and DFT persists up to 2~eV is therefore a strong indication that the low energy spectral weight is suppressed by the Hubbard $U$ and Hund's $J$-interaction.

\par 

\begin{figure}[t!!]
\begin{center}
\includegraphics[width=\columnwidth]{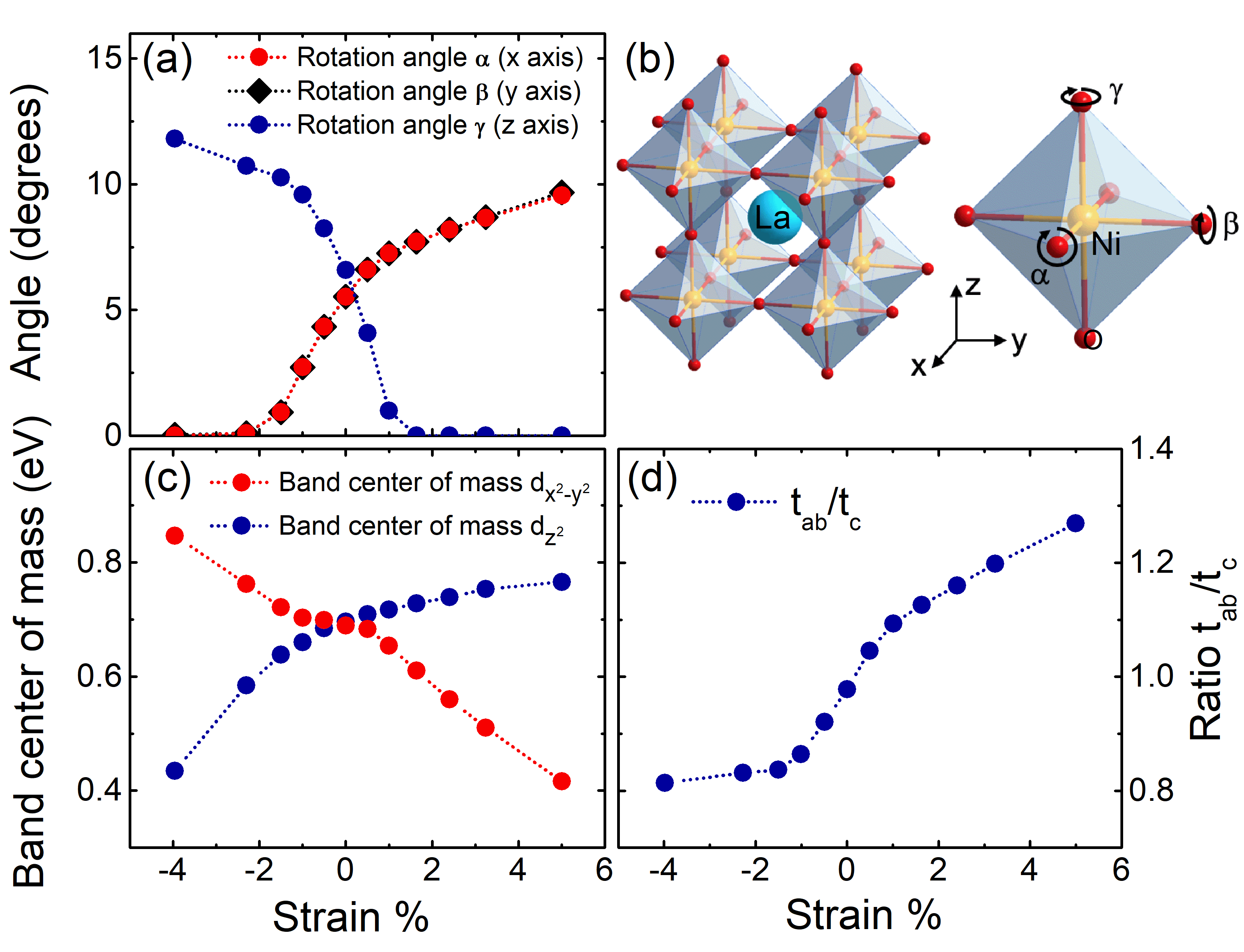}
\caption{\label{fig:electronic_structure} 
{\bf a}, Rotation of the oxygen octahedra along the $x$ and $y$ axis (corresponding to out-of-plane tilts $\alpha$ and $\beta$) and along the $z$ axis (corresponding to the in-plane rotation angle $\gamma$) extracted from the relaxed DFT crystal structures. Note that the values of $\alpha$ and $\beta$ are very close.
{\bf b}, Unit cell of LaNiO$_3$ and illustration of the tilt axes.
{\bf c}, Evolution of the intra-$e_g$ crystal field splitting, evaluated from the projected DFT density of states as a function of strain.
{\bf d}, Calculated ratio of the in-plane to out-of plane hopping amplitudes as a function of strain (see Appendix~\ref{appendix:theory}).} 
\end{center}
\end{figure}

\begin{figure*}[!ht]
\begin{center}
\includegraphics[width=2\columnwidth]{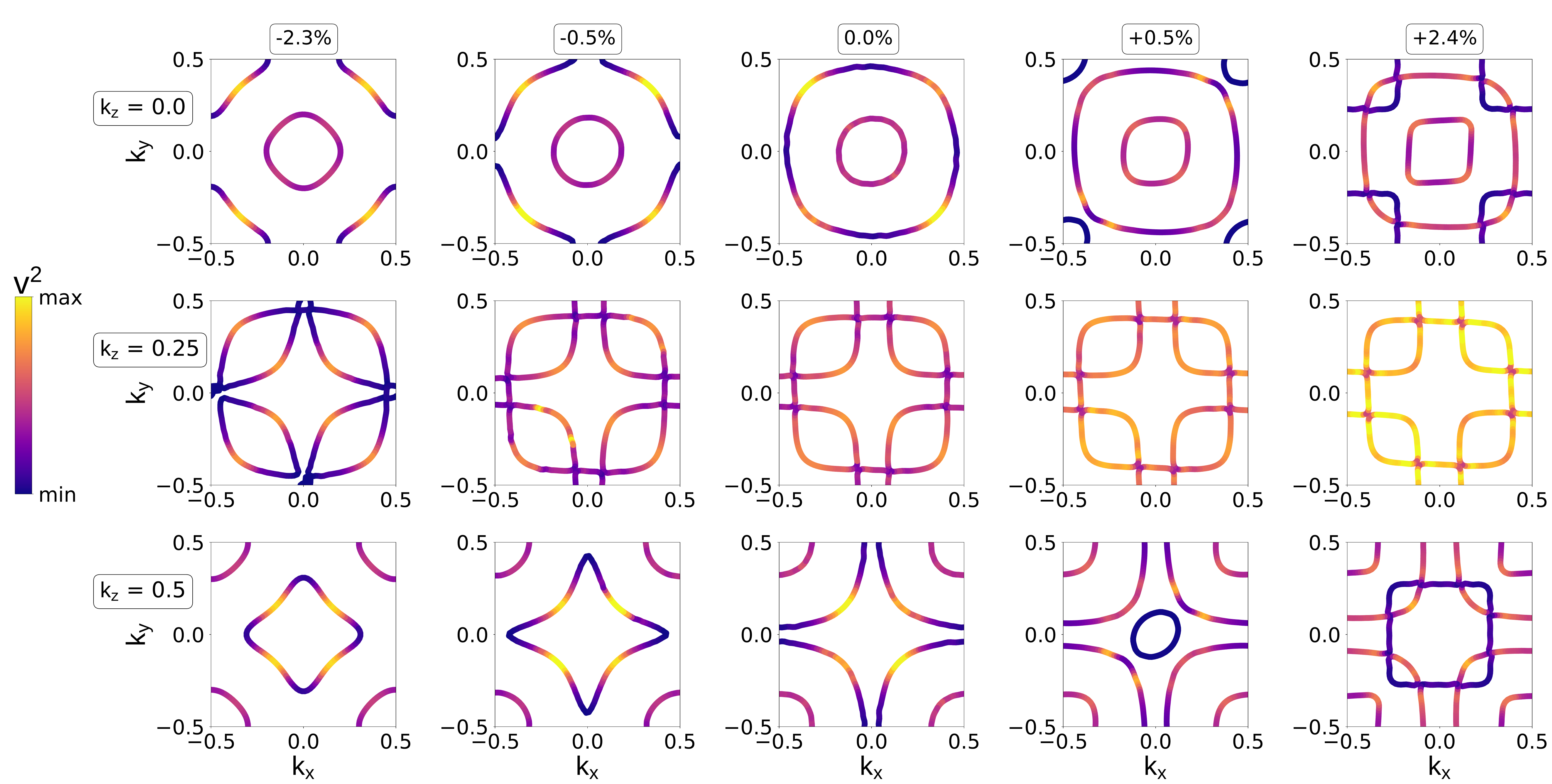}
\caption{
\label{fig:Fermi_surfaces}
DFT Fermi surface for 5 values of strain (columns). Each row corresponds to a different value of the out-of-plane momentum $k_z$ (in units of $\pi / c$).
The Lifshitz transition is manifest when a small compressive (see $k_z = 0.5$) or tensile (see $k_z = 0.0$) strain is applied.  
Along each Fermi surface sheet, the magnitude of the in-plane Fermi velocity $v_x^2+v_y^2$ is color-coded, demonstrating  the 
overall increase of the in-plane velocity under tensile strain.}
\end{center}
\end{figure*}

In order to understand the evolution of the electronic structure as a function of strain, we now turn to an in-depth analysis of how the structural changes evolve as a function of strain and, in turn, how they affect electronic structure as computed from DFT -- this is the focus of Figs.~\ref{fig:electronic_structure} and~\ref{fig:Fermi_surfaces}.
The first important consideration is a structural one. Fig.~\ref{fig:electronic_structure}{\bf a} displays the strain dependence of the rotations of the NiO$_6$ octahedra with respect to the ($x,y$) axis (`tilts') and with respect to the $z$-axis (`rotations'), as obtained from our DFT structural relaxation calculations (see Appendix~\ref{appendix:theory} for details). 
As reported in previous work~\cite{may2010,peil2014}, tensile strain promotes tilts and suppresses rotations, while compressive strain has the opposite effect. 
As a consequence, tensile strain distorts the out-of-plane Ni-O-Ni bond, keeping the in-plane bond angles almost unchanged \cite{fowlie2019}, 
making the (oxygen mediated) effective nickel-nickel in-plane hopping more favorable than the out-of-plane hopping.
Fig.~\ref{fig:electronic_structure}{\bf d} displays the ratio of in-plane to out-of-plane Ni-Ni effective hopping $t_{ab}/t_c$, obtained from a maximally localized Wannier function construction~\cite{MLWF1, MLWF2, wannier90} for Ni-centered orbitals of $e_g$ symmetry (see Appendix~\ref{appendix:theory}). This ratio is seen to increase from about 0.8 to 1.2 when going from compressive to tensile strain.
At the same time, the compression of the c-axis under tensile strain leads to an energetic destabilization of the out-of-plane antibonding $d_{z^2}$ orbital in comparison to the in-plane $d_{x^2-y^2}$ one. 

Fig.~\ref{fig:electronic_structure}{\bf c} displays the intra-$e_g$ crystal field splitting $\Delta_c \equiv \varepsilon(x^2-y^2)-\varepsilon(z^2)$ 
(relative stabilization energy of the in-plane orbital), defined from the center of mass of each band (the first moment), which is negative for tensile strain and positive for compressive strain.  
Both effects, the favoring of the in-plane hopping and the stabilization of the $d_{x^2-y^2}$ orbital, give rise to an increase of the conductivity when going from compressive to tensile strain. 
These considerations provide an insight into the important factors to explain the behavior of the conductivity. 
However, the low-frequency (Drude) part of the spectral weight is essentially a Fermi surface property, and the observed changes in the optical response therefore motivate a detailed analysis of the DFT Fermi surface.

In Fig.~\ref{fig:Fermi_surfaces} we display the calculated Fermi surfaces for five values of strain, as in-plane $(k_x, k_y)$ cuts for three different values of the out-of-plane momentum $k_z$.
The striking observation is that a transition in the topology of the Fermi surface (Lifshitz transition) is found as soon as a small strain is applied to the unstrained compound, for both compressive and tensile strain.
Focusing for example on the $k_z=0$ cut, we see that an additional Fermi surface sheet appears at the zone corner when a small tensile strain is applied, while the outer Fermi surface sheet switches from closed to open (i.e from $\Gamma$-centered to $X$-centered) under compressive strain. 
We have calculated the magnitude of the in-plane electronic velocity $v_x^2(\vec{k})+v_y^2(\vec{k})$ as the momentum $\vec{k}$ varies along each Fermi surface sheet, and displayed the result in Fig.~\ref{fig:Fermi_surfaces} as a color intensity map.
This reveals the considerable increase of the in-plane velocity from compressive to tensile strain, especially visible e.g. for the $k_z=0.25~\pi/c$ cut.

Overall, our calculated Fermi surfaces are in good agreement with the ARPES data of Yoo {\em et al.}~\cite{Yoo2015}. 
Comparing the images in panels (c) and (d) of Fig.~3 from Ref.~\onlinecite{Yoo2015} (corresponding to compressive and tensile strains, respectively), one can clearly see at least one signature of the Lifshitz transition. 
Specifically, on going from compressive to tensile strain, a pocket at point $M$ (folded to point $Z$) emerges. This is exactly the picture one can infer from our Fig.~\ref{fig:Fermi_surfaces} ($k_z = 0.5~\pi/c$ cut, strains from -2.3~\% to +2.4~\%). 
Another notable feature is a star-shaped sheet visible both in our Fig.~\ref{fig:Fermi_surfaces} ($k_z = 0.5~\pi/c$ cut, strains -0.5~\% and -2.3~\%) and in Fig.~3(c) of Ref.~\onlinecite{Yoo2015} displaying the $ZRA$ plane. 
It seems to us that the precise topology of this sheet is harder to determine unambiguously from the reported ARPES data. 
This sheet is definitely open at tensile strains and its neck is defined by a flat band close to the Fermi level (the band along $ZR$ in Fig.~1 of Ref.~\onlinecite{Yoo2015} and the $d_{z^2}$-like band along $\Gamma{}M$ in our Fig.~\ref{fig:DFT_DOS}). 
Both in our DFT results and in ARPES the band moves up when going from tensile to compressive strain. 
However, while in our calculations it clearly crosses the Fermi level at a strain slightly below zero, thus closing the star-shaped sheet, its fate is less clear in ARPES data at -1.3~\% strain. 
Yoo {\em et al.} infer the band position from the maxima of the energy-distribution curves (EDCs) and the results shown in Fig.~1 of Ref.~\onlinecite{Yoo2015} indicate that at -1.3~\% the $ZR$-band is still below the Fermi level. 
However, the band is very flat and lies very close to the Fermi level, and a more detailed analysis of the data may be necessary.  
Nevertheless, the comparison of trends observed in our calculations and in ARPES suggests that the star-shaped sheet closes at some value of the strain on the compressive side, even if slightly below -1.3~\% strain. 
Hence we conclude that there is overall reasonable qualitative agreement between our results and the ARPES data and a clear indication of a Fermi surface Lifshitz transition.

The strain dependence of the two key electronic structure parameters $\Delta_c$ and $t_{ab}/t_c$ can be directly connected to the evolution of the Fermi surface topology. 
We show in Appendix~\ref{appendix:theory} that by starting from an $e_g$ Wannier model for the (hypothetical) $P 4/mmm$ tetragonal structure without octahedral rotations and tilts, the calculated DFT trends in the Fermi surface can be semi-quantitatively explained when these two parameters are varied. 
We encourage the reader to explore this further with an interactive applet that we provide online~\cite{Binder}.
We emphasize that the evolution of the hopping anisotropy ($t_{ab} / t_c$) and the change in the crystal-field splitting ($\Delta_c$) both  result, when taken into account separately, in a decrease of the Drude weight when strain is increased. 
However, the magnitude of the changes seen in the conductivity (factor of two), as well as the evolution of the Fermi surface, can only be understood when both ingredients are considered together. 

Taken together, our findings on the calculated electronic structure provide a consistent explanation of the overall trends for the strain dependence of the optical spectra, and in particular explain why tensile strain leads to an increase of the Drude weight and optical conductivity. We note that a Lifshitz Fermi surface transition under strain has indeed been observed by ARPES spectroscopy for this material~\cite{Yoo2015}.
\section{Conclusions}
In conclusion, our optical measurements reveal a remarkable sensitivity of the electronic structure of LaNiO$_3$ to strain. 
In contrast to naive intuition, the in-plane low-frequency Drude weight increases by a factor close to two when we lattice-tune the material from highly compressive (-3.34\%) to moderate tensile (+1.75\%) biaxial strain. 
Our density functional theory calculations reveal that these effects are due to  drastic changes of the electronic structure under strain, caused by the changes in tilts and rotations of the NiO$_6$ octahedra. 
Our DFT calculations reveal a topological change of the Fermi surface (Lifshitz transition), which strongly modifies the velocity of carriers on the different Fermi surface sheets. 
We have provided a simple explanation of the evolution of the topology of the Fermi surface and of these trends, based on the changes of both the crystal field splitting and ratio of in-plane to out-of-plane hopping under strain. 
A prediction of our theory is that the Drude weight associated with the out-of-plane conductivity should correspondingly decrease from compressive to tensile strain.  
Our experimental results also reveal that interaction effects are sensitive to strain, with an enhancement factor of the optical effective mass ranging between about $1.5$ and $3.0$.
The evolution of the electronic structure ultimately results from the structural changes under strain (rotations, tilts and $c/a$ aspect ratio of the octahedra), illustrating the rich interplay between structural and electronic aspects of transition-metal oxides. 
This interplay can be leveraged in tuning electronic functionalities of oxides by strain engineering. 
Strain tuning of the electronic structure of LaNiO$_3$ and the detailed theoretical description thereof provide important tools for extending the use of this material in applications such as electrodes, gas-sensing and catalysis~\cite{catalano2018}.
\section{Acknowledgements.}
This project was supported by the Swiss National Science Foundation project 200020-179157 (DvdM), 200021-163103 (ABK), 200020-185061 (ABK),
200020-179155 (JMT), NCCR MARVEL (HS, OP), 
the European Research Council project 319286-QMAC (AG, JMT), 
and the \"Osterreichische Forschungsf\"orderungsgesellschaft COMET program IC-MPPE project 859480 (OP).
The Flatiron Institute is a division of the Simons Foundation.
NB acknowledges P. Marsik for useful discussions regarding the THz ellipsometry setup.
\appendix
\section{Thin film growth and characterization}
\label{appendix:films}
\begin{table*}[t!!]
\center
\begin{tabular}{|c|c|c|c|c|c|}
\hline
Substrate  & YAlO$_3$ & NdAlO$_3$ & LaAlO$_3$ & NdGaO$_3$ & SrTiO$_3$ \\
\hline
Space group &Pnma&R$\overline{3}$c&R$\overline{3}$c&Pnma&Pm$\overline{3}$m \\
\hline
Substrate $a_{sub,exp}$ $^{(i)}$ (\AA) & 3.71 & 3.74 & 3.789& 3.864 & 3.905 \\
\hline
Film $c$-axis constant $^{(ii)}$   (\AA) & 3.87 &   3.953 &   3.91 &  3.822 &  3.818 \\
\hline
Film cell volume    (\AA$^3$) & 53.3  & 55.3 & 56.1  & 57.1  & 58.2  \\               
\hline
Film strain & $-0.0334$ & $-0.0255$ & $-0.0128$& $+0.0068$ & $+0.0175$ \\   
\hline
Film thickness (uc)$^{(iii)}$ &  16  & 13  & 14  & 14  & 13 \\       
\hline
\end{tabular}
\caption{\label{table:samples}
Film parameters determined at room temperature. Pnma, R$\overline{3}$c and Pm$\overline{3}$m belong to the orthorhombic, rhombohedral and cubic crystal systems respectively.
{\em (i)} These values represent the geometric mean of the pseudocubic lattice constants parallel to the (001)-oriented substrate surfaces, and were obtained from out-of-plane X-ray diffraction data of the substrate measured at room temperature.
{\em (ii)} Pseudocubic $c$-axis parameter calculated from the $\theta \text{ - } 2\theta$ X-ray diffraction of the films. 
{\em (iii)} Thickness in units of the pseudocubic $c$-axis parameter, determined from the  $\theta \text{ - } 2\theta$ X-ray diffraction pattern of the films.}
\end{table*}
We used commercial single-crystal YAlO$_3$, NdAlO$_3$, LaAlO$_3$, NdGaO$_3$ and SrTiO$_3$ substrates from CrysTec GmbH with a $5\times5$~\SI{}{mm^2} surface area and \SI{0.5}{mm} thickness. 
All surfaces were (001)-oriented in pseudo-cubic notation. Prior to deposition all substrates were thermally treated in flowing oxygen to ensure atomically-flat step-terrace surfaces. 
For each film/substrate a combination of two identical substrates was used at the same time to produce twin samples, one for optics and the other for DC transport measurements. 
The epitaxial LaNiO$_3$ films were deposited via radio frequency off-axis magnetron sputtering at a temperature of \SI{510}{\celsius} in an oxygen:argon mix of 2:7 maintained at \SI{0.24}{mbar}.

After deposition, the surface quality of the films was verified by atomic force microscopy and the step-terrace topography of the substrate was seen to be retained. 
The crystalline properties were checked using X-ray diffraction. 
Bragg-Brentano scans showed finite thickness oscillations, indicating a high quality film and providing the precise film thickness after fitting~\cite{Lichtensteiger2018} (see Table~\ref{table:samples}). 
The positions of the peaks gave the lattice constants of each film/substrate. 

\begin{figure}[!h]
\begin{center}
\includegraphics[width=\columnwidth]{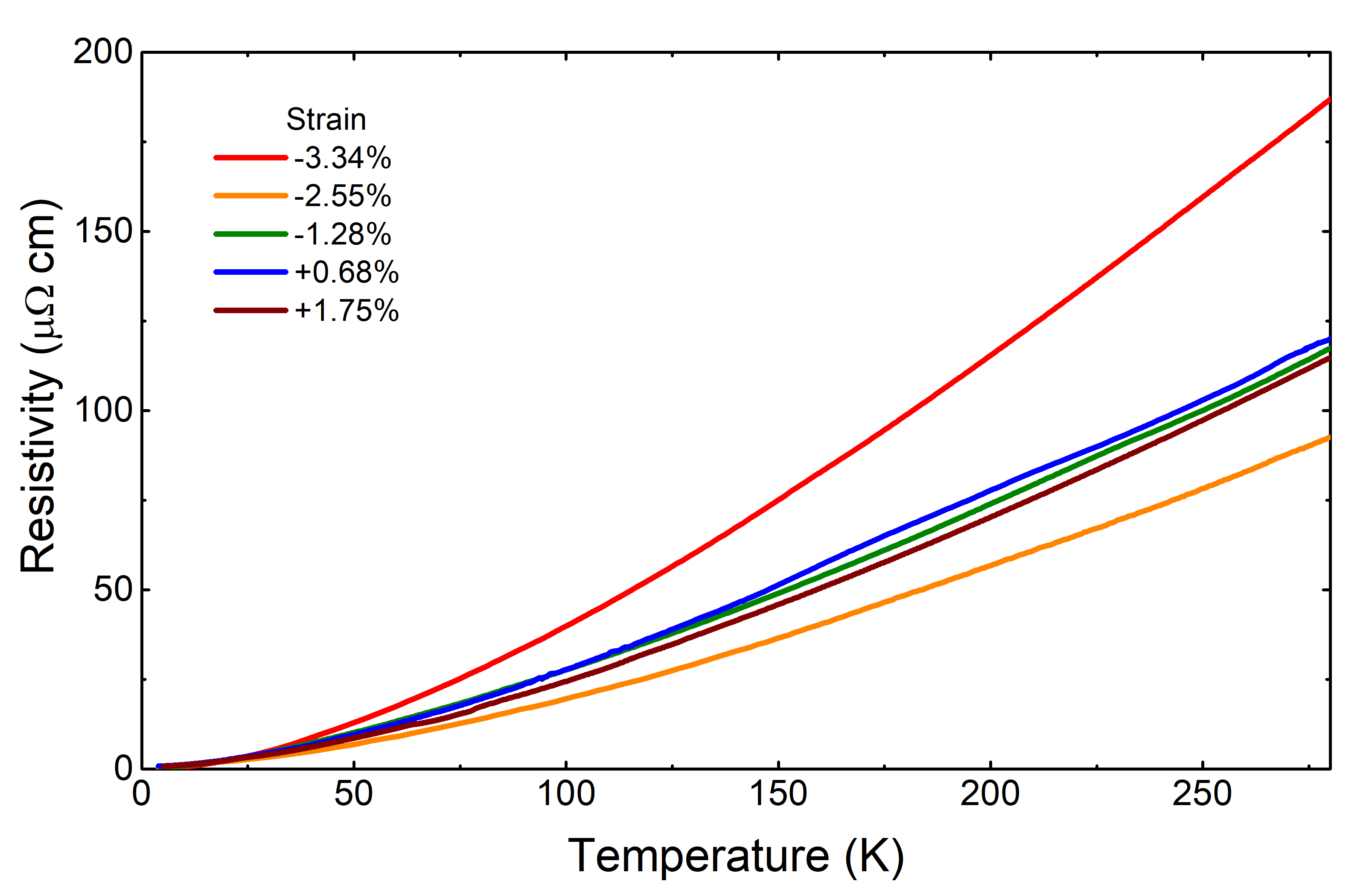}
\caption{\label{fig:rho_DC} 
Temperature dependent part of the DC resistivity, $\rho(T)-\rho_{0}$ of the five LaNiO$_3$ film/substrate combinations YAlO$_3$ (-3.34\%), NdAlO$_3$ (-2.55\%), LaAlO$_3$ (-1.28\%), NdGaO$_3$ (+0.68\%), SrTiO$_3$ (+1.75\%), where $\rho_{0}$ is the residual resistivity. The values of $\rho_{0}$ are 167, 48, 48, 68, and 50~$\mu\Omega$cm for YAlO$_3$, NdAlO$_3$, LaAlO$_3$, NdGaO$_3$, and SrTiO$_3$ respectively.}
\end{center}
\end{figure}

For each film/substrate combination the member of the twin samples destined for transport measurements was etched into multiple $100 \times 680$~\SI{}{\micro m} Hall bars, contacted with sputtered platinum contacts and bonded with aluminum wires. 
Then the DC resistivity (displayed in Fig.~\ref{fig:rho_DC}) was recorded in a $^4$He dipping station down to~\SI{4}{K}. 

The presence of twin domains in the LaAlO$ _3$ and NdAlO$_3$ substrates influences the terahertz transmission. 
In the case of LaAlO$_3$ and LaNiO$_3$ films on LaAlO$_3$ substrates, we managed to cancel this interference effect by a proper incoming polarisation orientation. 
This was not possible for NdAlO$_3$ and LaNiO$_3$ films on NdAlO$_3$ substrates. 
\section{Electronic band structure calculations}
\label{appendix:theory}
\begin{figure*}[!ht]
\begin{center}
\includegraphics[width=2\columnwidth]{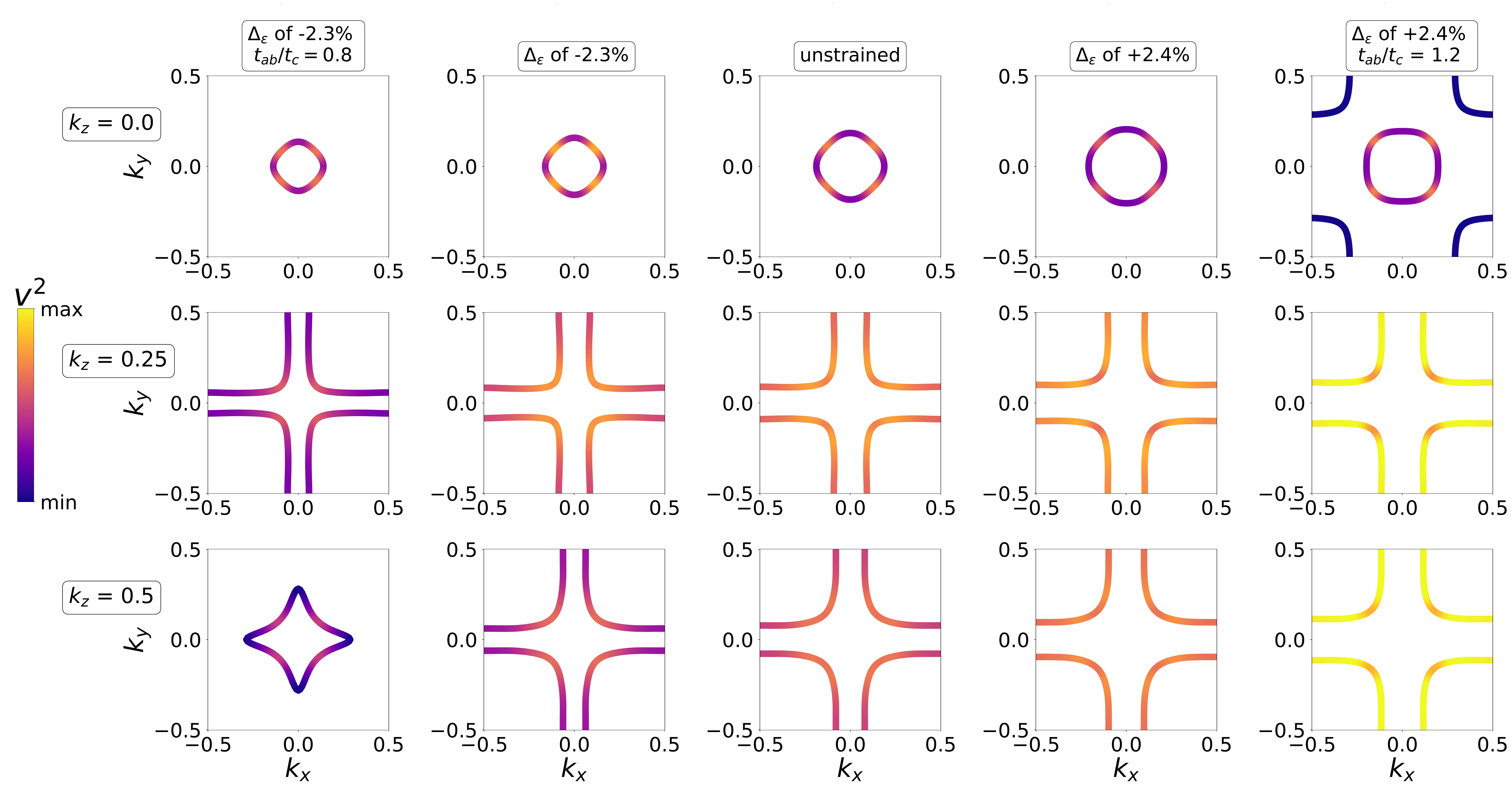}
\caption{\label{fig:Fermi_tetragonal} 
Fermi surface of the unstrained tetragonal structure without octahedral rotations and tilts calculated with DFT and a subsequent Wannier model construction (middle column). The second and fourth columns show the Fermi surface of the same model, but with the crystal-field splitting $\Delta_c$ of -2.3\% and +2.4\% tetragonal structures, respectively. The first and last columns show the effect of scaling the hopping ratio $t_{ab}/t_{c}$ by 0.8 and 1.2 in addition to the modified $\Delta_c$. Each row corresponds to a different value of the out-of-plane momentum $k_z$ (in units of $\pi / c$). Along each Fermi surface sheet the magnitude of the in-plane Fermi velocity $v_x^2+v_y^2$ is color-coded. Note that the additional sheets visible in Fig.~\ref{fig:Fermi_surfaces} are Fermi surface reconstructions due to a two Ni-atom (twice as large) unit cell.}
\end{center}
\end{figure*}

Crystal structure relaxation was performed within the generalized gradient approximation (Perdew-Burke-Ernzerhof parametrization, PBE)~\cite{pbe96} using the projected-augmented waves (PAW) method~\cite{blochl1994} as implemented in the Vienna Ab-initio Simulation Package (VASP)~\cite{paw_vasp,vasp1,vasp2}.
The integration over the Brillouin zone was done using a $k$-mesh with $11\times 11\times 11$ points and a plane-wave cutoff of $E_{\textrm{cut}} = \SI{600}{eV}$. 
Structure relaxation was considered converged when the forces were smaller than \SI{e-3}{eV\per \angstrom}.
The in-plane velocities of the states on the Fermi surfaces, shown in Fig.~\ref{fig:Fermi_surfaces} and Fig.~\ref{fig:Fermi_tetragonal}  have been calculated with the BoltzTraP2 software package~\cite{madsen2018}.

A film of LaNiO$_{3}$ was simulated using a $C2/c$ unit cell, which corresponds to $a^{-}a^{-}c^{-}$ octahedral rotation pattern in Glazer notation~\cite{glazer1972}.
The primitive cell of this structure contains four formula units with oppositely oriented tilts and rotations of the NiO$_6$ octahedra.
The strain effect of the substrate was taken into account by constraining the in-plane pseudo-cubic vectors $\mathbf{a}_{p}$, $\mathbf{b}_{p}$ to the corresponding vectors of a generic cubic substrate. 
The strain values are defined with respect to the lattice constant of the PBE-relaxed bulk structure ($a_{p} \approx \SI{3.863}{\angstrom}$).
We used the method of Ref.~\onlinecite{peil2014}. The in-plane lattice parameters were fixed and all other degrees of freedom, such as the out-of-plane lattice vector, oxygen, and cation positions, are allowed to relax.
For most of the strains, except for the very large ones, the final space group of the relaxed unit cell remained $C2/c$.
For the extreme compressive strain (-4.0\%)  with vanishing $\mathbf{a}$-, $\mathbf{b}$-rotation angles, $\alpha = \beta = 0\si{\degree}$, and for the extreme tensile strain (+5.0\%) with $\gamma=0\si{\degree}$ the relaxation converged to unit cells corresponding to supergroups of $C2/c$: $I4/mcm$ under compression and $Fmmm$ under tension.

\begin{figure}[!ht]
\begin{center}
\includegraphics[width=\columnwidth]{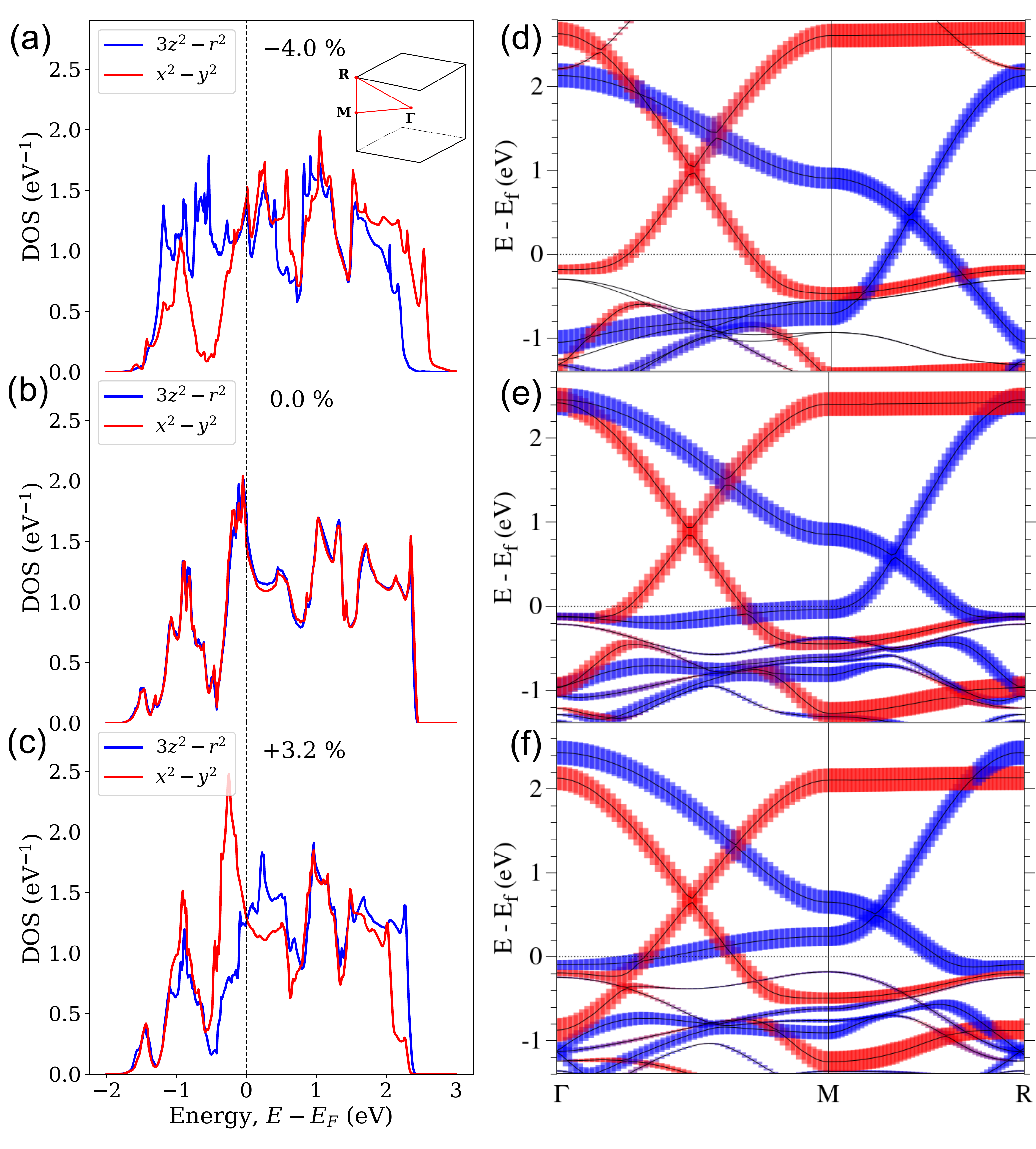}
\caption{\label{fig:DFT_DOS} 
{\bf a-c}, DFT density of states projected on $d_{z^2}$ (blue) and $d_{x^2-y^2}$ (red) orbitals. {\bf d-f} energy momentum dispersion along the trajectory shown in the inset of {\bf a}, with orbital character for three different strain values: -4.0\% ({\bf a, d}), 0.0\% ({\bf b, e}), +3.2\% ({\bf c, f}). 
}
\end{center}
\end{figure}
{In Fig.~\ref{fig:DFT_DOS} we show the resulting density of states and the band structure along a high symmetry path through the Brillouin zone for three different strains.} 
To evaluate the orbital character of the band structure, as shown in Fig.~\ref{fig:DFT_DOS}{\bf d-f}, we projected the Kohn-Sham wavefunctions onto $e_{g}$-like ($d_{z^{2}}$, $d_{x^{2}-y^{2}}$) local states centered on the Ni atoms. We selected band states in a narrow energy window ([\num{-1.6}, \num{4.0}]~\SI{}{eV}) around the Fermi level, corresponding to  antibonding states in the local frame of octahedra, which are then transformed into projected localized orbitals (PLOs) as defined in Refs.~\onlinecite{amadon2008,schuler2018}.

To calculate the hopping ratios $t_{ab} / t_{c}$, shown in Fig.~\ref{fig:electronic_structure}, we constructed a low-energy Wannier Hamiltonian on a $10\times 10\times 10$ k-grid using maximally-localized Wannier functions~\cite{MLWF1,MLWF2,wannier90} of Ni-$e_g$ symmetry. For all strains we used a frozen energy window ([\num{0.04}, \num{0.74}]~\SI{}{eV}) to improve the quality of the Wannier fit. 
We checked that the Wannier functions are centered exactly on the atomic positions and confirmed that the resulting Wannier Hamiltonian is real. 
However, it should be noted that the employed $e_g$-like basis is not sufficient to accurately match the band structure for states below the frozen energy. 
From the Wannier model we extracted the on-site energies (Fig.~\ref{fig:tau_strain}{\bf c}), which show the same trend as the band center of mass shown in main text Fig.~\ref{fig:electronic_structure} {\bf c}. 
Additionally, we calculated effective in-plane $t_{ab}$ and out-of-plane $t_{c}$ hopping amplitudes (Fig.~\ref{fig:tau_strain}{\bf b}) via
\begin{eqnarray}
    t_{ab} &=& \sum_{\mathbf{R}\in\{ab\}} \frac{\sqrt{\sum_{i,j} |t_{i,j}(\mathbf{R})|^2}}{N_{\mathbf{|R|}}} 
    \nonumber \\ 
     t_{c} &=& \sum_{\mathbf{R}\in\{c\}} \frac{\sqrt{\sum_{i,j} |t_{i,j}(\mathbf{R})|^2}}{N_{\mathbf{|R|}}}
\label{Eq:hopping}
\end{eqnarray}
{For $t_{ab}$ the sum runs over all real-space vectors $\mathbf{R}\in\{ab\}$ without any out-of-plane component and without any in-plane components for $t_{c}$, respectively.} 
The indices $i,j \in \{z^2, x^2-y^2\}$ are orbital indices and $N_\mathbf{|R|}$ is the number of Ni-atoms at the same distance $\mathbf{|R|}$, {\em e.g.} there are two nearest neighbor out-of-plane hoppings and four nearest neighbor in-plane hoppings.

Optical conductivities for the relaxed structures have been calculated using WIEN2k~\cite{Blaha2018} and the transport code implemented in the TRIQS/DFTTools package~\cite{TRIQS/DFTTOOLS}, which is based on the TRIQS library~\cite{TRIQS}. 
We used a denser $k$-mesh of $33\times 33\times 33$ points and assumed a fixed scattering rate of $\hbar / \tau = \SI{60}{meV}$ for all strains. 
The DC conductivity was calculated from the relation $\sigma_{DC}=\tau\omega_{p}^2/(4\pi)$, where $\omega_{p}^2$ is the spectral weight tensor defined as
\begin{equation}
\omega_{p}^2=   
\frac{4 \pi e^2}{\Omega_{pc}}
\sum\limits_{\vec{k},j}^{{1^t}BZ} 
{\vec{v}} _{j}(\vec{k})
{\vec{v}} _{j}(\vec{k})
\left(
-\frac{\partial f_{\vec{k},j}}{\partial \epsilon_{\vec{k},j}}
\right),
\end{equation}
${\vec{v}} _{j}(\vec{k})$ is the group velocity of the $j$'th band for momentum value $\vec{k}$, $\Omega_{pc}$ is the volume of the primitive cell, and $f_{\vec{k},j}=1/[1+\exp{(\epsilon_{\vec{k},j}/k_BT)}]$ is the Fermi-Dirac distribution. 
As discussed in the main text, under tensile strain the in-plane DC conductivity increases. 
In the calculations, the out-of-plane DC conductivity shows the exact opposite behavior, it decreases under tensile strain (see Fig.~\ref{fig:tau_strain}{\bf a})

To demonstrate the effect of oxygen-octahedra rotations and tilts we also performed calculations for (hypothetical) undistorted $P 4/mmm$ tetragonal structures using the in- and out-of-plane lattice constants of the fully relaxed distorted structures. 
The octahedral distortions, rotations, and tilts allow for optical inter-band transitions, leading to a pronounced weight in the optical conductivity around $\sim \SI{1}{eV}$ observed in theory and experiment (see main text Fig.~\ref{fig:sigma}). 
This weight is not present in the optical conductivity of the undistorted tetragonal structures, as shown in Fig.~\ref{fig:DFT_sigma}. 
In the low-frequency region of the optical conductivity we observe two effects: First, the DC conductivities are larger for the tetragonal structures, due to larger inter-site hoppings as a result of straight bonds. 
Second, the trend of a conductivity decrease under increased strain is still present (although only weakly pronounced for tensile strains), which can be traced back to how the crystal-field splitting evolves under strain (see Fig.~\ref{fig:neff}).

In Fig.~\ref{fig:Fermi_tetragonal}  we show the effect on the Fermi surface and the velocities when tuning the crystal field splitting $\Delta_c$ and the hopping ratio $t_{ab}/t_{c}$ starting from the unstrained tetragonal structure (middle column). 
Note that the additional sheets visible in Fig.~\ref{fig:Fermi_surfaces} of the main text are Fermi surface reconstructions due to a two Ni-atom (twice as large) unit cell. 
Only if both the change in $\Delta_c$ and the change in $t_{ab}/t_{c}$ are accounted for (first and last column) can the Lifshitz transitions under strain be qualitatively reproduced. 
The changes in $\Delta_c$ and $t_{ab}/t_{c}$ also lead to higher in-plane velocities, when adjusted in accordance with the effect of tensile strain (i.e. $\Delta_c$ decreases and $t_{ab}/t_{c}$ increases), as seen from the color-coding of the Fermi surfaces. 
We encourage the reader to further explore the effect of $\Delta_c$ and $t_{ab}/t_{c}$  on the Fermi surface and the band structure with our interactive online applet~(Ref.~\onlinecite{Binder}).

\section{Data availability}
The datasets generated and analyzed during the current study are available in Ref.~\onlinecite{yareta}. These will be preserved for 10 years. All other data that support the plots within this paper and other findings of this study are available from the corresponding author upon reasonable request.
\vspace{20\baselineskip}
%

\begin{thebibliography}{65}%
\makeatletter
\providecommand \@ifxundefined [1]{%
 \@ifx{#1\undefined}
}%
\providecommand \@ifnum [1]{%
 \ifnum #1\expandafter \@firstoftwo
 \else \expandafter \@secondoftwo
 \fi
}%
\providecommand \@ifx [1]{%
 \ifx #1\expandafter \@firstoftwo
 \else \expandafter \@secondoftwo
 \fi
}%
\providecommand \natexlab [1]{#1}%
\providecommand \enquote  [1]{``#1''}%
\providecommand \bibnamefont  [1]{#1}%
\providecommand \bibfnamefont [1]{#1}%
\providecommand \citenamefont [1]{#1}%
\providecommand \href@noop [0]{\@secondoftwo}%
\providecommand \href [0]{\begingroup \@sanitize@url \@href}%
\providecommand \@href[1]{\@@startlink{#1}\@@href}%
\providecommand \@@href[1]{\endgroup#1\@@endlink}%
\providecommand \@sanitize@url [0]{\catcode `\\12\catcode `\$12\catcode
  `\&12\catcode `\#12\catcode `\^12\catcode `\_12\catcode `\%12\relax}%
\providecommand \@@startlink[1]{}%
\providecommand \@@endlink[0]{}%
\providecommand \url  [0]{\begingroup\@sanitize@url \@url }%
\providecommand \@url [1]{\endgroup\@href {#1}{\urlprefix }}%
\providecommand \urlprefix  [0]{URL }%
\providecommand \Eprint [0]{\href }%
\providecommand \doibase [0]{http://dx.doi.org/}%
\providecommand \selectlanguage [0]{\@gobble}%
\providecommand \bibinfo  [0]{\@secondoftwo}%
\providecommand \bibfield  [0]{\@secondoftwo}%
\providecommand \translation [1]{[#1]}%
\providecommand \BibitemOpen [0]{}%
\providecommand \bibitemStop [0]{}%
\providecommand \bibitemNoStop [0]{.\EOS\space}%
\providecommand \EOS [0]{\spacefactor3000\relax}%
\providecommand \BibitemShut  [1]{\csname bibitem#1\endcsname}%
\let\auto@bib@innerbib\@empty
\bibitem [{\citenamefont {Torrance}\ \emph {et~al.}(1992)\citenamefont
  {Torrance}, \citenamefont {Lacorre}, \citenamefont {Nazzal}, \citenamefont
  {Ansaldo},\ and\ \citenamefont {Niedermayer}}]{torrance1992}%
  \BibitemOpen
  \bibfield  {author} {\bibinfo {author} {\bibfnamefont {J.~B.}\ \bibnamefont
  {Torrance}}, \bibinfo {author} {\bibfnamefont {P.}~\bibnamefont {Lacorre}},
  \bibinfo {author} {\bibfnamefont {A.~I.}\ \bibnamefont {Nazzal}}, \bibinfo
  {author} {\bibfnamefont {E.~J.}\ \bibnamefont {Ansaldo}}, \ and\ \bibinfo
  {author} {\bibfnamefont {C.}~\bibnamefont {Niedermayer}},\ }\href {\doibase
  10.1103/PhysRevB.45.8209} {\bibfield  {journal} {\bibinfo  {journal} {Phys.
  Rev. B}\ }\textbf {\bibinfo {volume} {45}},\ \bibinfo {pages} {8209}
  (\bibinfo {year} {1992})}\BibitemShut {NoStop}%
\bibitem [{\citenamefont {Garc\'{\i}a-Mu\~noz}\ \emph
  {et~al.}(1992)\citenamefont {Garc\'{\i}a-Mu\~noz}, \citenamefont
  {Rodr\'{\i}guez-Carvajal}, \citenamefont {Lacorre},\ and\ \citenamefont
  {Torrance}}]{garcia1992}%
  \BibitemOpen
  \bibfield  {author} {\bibinfo {author} {\bibfnamefont {J.~L.}\ \bibnamefont
  {Garc\'{\i}a-Mu\~noz}}, \bibinfo {author} {\bibfnamefont {J.}~\bibnamefont
  {Rodr\'{\i}guez-Carvajal}}, \bibinfo {author} {\bibfnamefont
  {P.}~\bibnamefont {Lacorre}}, \ and\ \bibinfo {author} {\bibfnamefont
  {J.~B.}\ \bibnamefont {Torrance}},\ }\href {\doibase
  10.1103/PhysRevB.46.4414} {\bibfield  {journal} {\bibinfo  {journal} {Phys.
  Rev. B}\ }\textbf {\bibinfo {volume} {46}},\ \bibinfo {pages} {4414}
  (\bibinfo {year} {1992})}\BibitemShut {NoStop}%
\bibitem [{\citenamefont {Catalan}(2008)}]{catalan2008}%
  \BibitemOpen
  \bibfield  {author} {\bibinfo {author} {\bibfnamefont {G.}~\bibnamefont
  {Catalan}},\ }\href {\doibase 10.1080/01411590801992463} {\bibfield
  {journal} {\bibinfo  {journal} {Phase Transitions}\ }\textbf {\bibinfo
  {volume} {81}},\ \bibinfo {pages} {729} (\bibinfo {year} {2008})}\BibitemShut
  {NoStop}%
\bibitem [{\citenamefont {Chaloupka}\ and\ \citenamefont
  {Khaliullin}(2008)}]{chaloupka2008}%
  \BibitemOpen
  \bibfield  {author} {\bibinfo {author} {\bibfnamefont {J.}~\bibnamefont
  {Chaloupka}}\ and\ \bibinfo {author} {\bibfnamefont {G.}~\bibnamefont
  {Khaliullin}},\ }\href {\doibase 10.1103/PhysRevLett.100.016404} {\bibfield
  {journal} {\bibinfo  {journal} {Phys. Rev. Lett.}\ }\textbf {\bibinfo
  {volume} {100}},\ \bibinfo {pages} {016404} (\bibinfo {year}
  {2008})}\BibitemShut {NoStop}%
\bibitem [{\citenamefont {Kumah}\ \emph {et~al.}(2014)\citenamefont {Kumah},
  \citenamefont {Disa}, \citenamefont {Ngai}, \citenamefont {Chen},
  \citenamefont {Malashevich}, \citenamefont {Reiner}, \citenamefont
  {Ismail-Beigi}, \citenamefont {Walker},\ and\ \citenamefont
  {Ahn}}]{LNOam2014}%
  \BibitemOpen
  \bibfield  {author} {\bibinfo {author} {\bibfnamefont {D.}~\bibnamefont
  {Kumah}}, \bibinfo {author} {\bibfnamefont {A.}~\bibnamefont {Disa}},
  \bibinfo {author} {\bibfnamefont {J.}~\bibnamefont {Ngai}}, \bibinfo {author}
  {\bibfnamefont {H.}~\bibnamefont {Chen}}, \bibinfo {author} {\bibfnamefont
  {A.}~\bibnamefont {Malashevich}}, \bibinfo {author} {\bibfnamefont {J.~W.}\
  \bibnamefont {Reiner}}, \bibinfo {author} {\bibfnamefont {S.}~\bibnamefont
  {Ismail-Beigi}}, \bibinfo {author} {\bibfnamefont {F.~J.}\ \bibnamefont
  {Walker}}, \ and\ \bibinfo {author} {\bibfnamefont {C.}~\bibnamefont {Ahn}},\
  }\href {\doibase 10.1002/adma.201304256} {\bibfield  {journal} {\bibinfo
  {journal} {Adv. Mater.}\ }\textbf {\bibinfo {volume} {26}},\ \bibinfo {pages}
  {1935} (\bibinfo {year} {2014})}\BibitemShut {NoStop}%
\bibitem [{\citenamefont {Ismail-Beigi}\ \emph {et~al.}(2017)\citenamefont
  {Ismail-Beigi}, \citenamefont {Walker}, \citenamefont {Disa}, \citenamefont
  {Rabe},\ and\ \citenamefont {Ahn}}]{Pico}%
  \BibitemOpen
  \bibfield  {author} {\bibinfo {author} {\bibfnamefont {S.}~\bibnamefont
  {Ismail-Beigi}}, \bibinfo {author} {\bibfnamefont {F.~J.}\ \bibnamefont
  {Walker}}, \bibinfo {author} {\bibfnamefont {A.~S.}\ \bibnamefont {Disa}},
  \bibinfo {author} {\bibfnamefont {K.~M.}\ \bibnamefont {Rabe}}, \ and\
  \bibinfo {author} {\bibfnamefont {C.}~\bibnamefont {Ahn}},\ }\href {\doibase
  10.1038/natrevmats.2017.60} {\bibfield  {journal} {\bibinfo  {journal} {Nat.
  Rev. Mater.}\ }\textbf {\bibinfo {volume} {2}},\ \bibinfo {pages} {17060}
  (\bibinfo {year} {2017})}\BibitemShut {NoStop}%
\bibitem [{\citenamefont {Phillips}\ \emph {et~al.}(2017)\citenamefont
  {Phillips}, \citenamefont {Rui}, \citenamefont {Georgescu}, \citenamefont
  {Disa}, \citenamefont {Longo}, \citenamefont {Okunishi}, \citenamefont
  {Walker}, \citenamefont {Ahn}, \citenamefont {Ismail-Beigi},\ and\
  \citenamefont {Klie}}]{Trilayer}%
  \BibitemOpen
  \bibfield  {author} {\bibinfo {author} {\bibfnamefont {P.~J.}\ \bibnamefont
  {Phillips}}, \bibinfo {author} {\bibfnamefont {X.}~\bibnamefont {Rui}},
  \bibinfo {author} {\bibfnamefont {A.~B.}\ \bibnamefont {Georgescu}}, \bibinfo
  {author} {\bibfnamefont {A.~S.}\ \bibnamefont {Disa}}, \bibinfo {author}
  {\bibfnamefont {P.}~\bibnamefont {Longo}}, \bibinfo {author} {\bibfnamefont
  {E.}~\bibnamefont {Okunishi}}, \bibinfo {author} {\bibfnamefont
  {F.}~\bibnamefont {Walker}}, \bibinfo {author} {\bibfnamefont {C.~H.}\
  \bibnamefont {Ahn}}, \bibinfo {author} {\bibfnamefont {S.}~\bibnamefont
  {Ismail-Beigi}}, \ and\ \bibinfo {author} {\bibfnamefont {R.~F.}\
  \bibnamefont {Klie}},\ }\href {\doibase 10.1103/PhysRevB.95.205131}
  {\bibfield  {journal} {\bibinfo  {journal} {Phys. Rev. B}\ }\textbf {\bibinfo
  {volume} {95}},\ \bibinfo {pages} {205131} (\bibinfo {year}
  {2017})}\BibitemShut {NoStop}%
\bibitem [{\citenamefont {Disa}\ \emph {et~al.}(2017)\citenamefont {Disa},
  \citenamefont {Georgescu}, \citenamefont {Hart}, \citenamefont {Kumah},
  \citenamefont {Shafer}, \citenamefont {Arenholz}, \citenamefont {Arena},
  \citenamefont {Ismail-Beigi}, \citenamefont {Taheri}, \citenamefont
  {Walker},\ and\ \citenamefont {Ahn}}]{PRMDisa}%
  \BibitemOpen
  \bibfield  {author} {\bibinfo {author} {\bibfnamefont {A.~S.}\ \bibnamefont
  {Disa}}, \bibinfo {author} {\bibfnamefont {A.~B.}\ \bibnamefont {Georgescu}},
  \bibinfo {author} {\bibfnamefont {J.~L.}\ \bibnamefont {Hart}}, \bibinfo
  {author} {\bibfnamefont {D.~P.}\ \bibnamefont {Kumah}}, \bibinfo {author}
  {\bibfnamefont {P.}~\bibnamefont {Shafer}}, \bibinfo {author} {\bibfnamefont
  {E.}~\bibnamefont {Arenholz}}, \bibinfo {author} {\bibfnamefont {D.~A.}\
  \bibnamefont {Arena}}, \bibinfo {author} {\bibfnamefont {S.}~\bibnamefont
  {Ismail-Beigi}}, \bibinfo {author} {\bibfnamefont {M.~L.}\ \bibnamefont
  {Taheri}}, \bibinfo {author} {\bibfnamefont {F.~J.}\ \bibnamefont {Walker}},
  \ and\ \bibinfo {author} {\bibfnamefont {C.~H.}\ \bibnamefont {Ahn}},\ }\href
  {\doibase 10.1103/PhysRevMaterials.1.024410} {\bibfield  {journal} {\bibinfo
  {journal} {Phys. Rev. Materials}\ }\textbf {\bibinfo {volume} {1}},\ \bibinfo
  {pages} {024410} (\bibinfo {year} {2017})}\BibitemShut {NoStop}%
\bibitem [{\citenamefont {Lee}\ \emph {et~al.}(2019)\citenamefont {Lee},
  \citenamefont {Lee}, \citenamefont {Georgescu}, \citenamefont {Fabbris},
  \citenamefont {Han}, \citenamefont {Zhu}, \citenamefont {Freeland},
  \citenamefont {Disa}, \citenamefont {Jia}, \citenamefont {Dean},
  \citenamefont {Walker}, \citenamefont {Ismail-Beigi},\ and\ \citenamefont
  {Ahn}}]{Cobaltate}%
  \BibitemOpen
  \bibfield  {author} {\bibinfo {author} {\bibfnamefont {S.}~\bibnamefont
  {Lee}}, \bibinfo {author} {\bibfnamefont {A.~T.}\ \bibnamefont {Lee}},
  \bibinfo {author} {\bibfnamefont {A.~B.}\ \bibnamefont {Georgescu}}, \bibinfo
  {author} {\bibfnamefont {G.}~\bibnamefont {Fabbris}}, \bibinfo {author}
  {\bibfnamefont {M.-G.}\ \bibnamefont {Han}}, \bibinfo {author} {\bibfnamefont
  {Y.}~\bibnamefont {Zhu}}, \bibinfo {author} {\bibfnamefont {J.~W.}\
  \bibnamefont {Freeland}}, \bibinfo {author} {\bibfnamefont {A.~S.}\
  \bibnamefont {Disa}}, \bibinfo {author} {\bibfnamefont {Y.}~\bibnamefont
  {Jia}}, \bibinfo {author} {\bibfnamefont {M.~P.~M.}\ \bibnamefont {Dean}},
  \bibinfo {author} {\bibfnamefont {F.~J.}\ \bibnamefont {Walker}}, \bibinfo
  {author} {\bibfnamefont {S.}~\bibnamefont {Ismail-Beigi}}, \ and\ \bibinfo
  {author} {\bibfnamefont {C.~H.}\ \bibnamefont {Ahn}},\ }\href {\doibase
  10.1103/PhysRevLett.123.117201} {\bibfield  {journal} {\bibinfo  {journal}
  {Phys. Rev. Lett.}\ }\textbf {\bibinfo {volume} {123}},\ \bibinfo {pages}
  {117201} (\bibinfo {year} {2019})}\BibitemShut {NoStop}%
\bibitem [{\citenamefont {Mizokawa}\ \emph {et~al.}(2000)\citenamefont
  {Mizokawa}, \citenamefont {Khomskii},\ and\ \citenamefont
  {Sawatzky}}]{mizokawa2000}%
  \BibitemOpen
  \bibfield  {author} {\bibinfo {author} {\bibfnamefont {T.}~\bibnamefont
  {Mizokawa}}, \bibinfo {author} {\bibfnamefont {D.~I.}\ \bibnamefont
  {Khomskii}}, \ and\ \bibinfo {author} {\bibfnamefont {G.~A.}\ \bibnamefont
  {Sawatzky}},\ }\href {\doibase 10.1103/PhysRevB.61.11263} {\bibfield
  {journal} {\bibinfo  {journal} {Phys. Rev. B}\ }\textbf {\bibinfo {volume}
  {61}},\ \bibinfo {pages} {11263} (\bibinfo {year} {2000})}\BibitemShut
  {NoStop}%
\bibitem [{\citenamefont {Park}\ \emph {et~al.}(2012)\citenamefont {Park},
  \citenamefont {Millis},\ and\ \citenamefont {Marianetti}}]{park2012}%
  \BibitemOpen
  \bibfield  {author} {\bibinfo {author} {\bibfnamefont {H.}~\bibnamefont
  {Park}}, \bibinfo {author} {\bibfnamefont {A.~J.}\ \bibnamefont {Millis}}, \
  and\ \bibinfo {author} {\bibfnamefont {C.~A.}\ \bibnamefont {Marianetti}},\
  }\href {\doibase 10.1103/PhysRevLett.109.156402} {\bibfield  {journal}
  {\bibinfo  {journal} {Phys. Rev. Lett.}\ }\textbf {\bibinfo {volume} {109}},\
  \bibinfo {pages} {156402} (\bibinfo {year} {2012})}\BibitemShut {NoStop}%
\bibitem [{\citenamefont {Mazin}\ \emph {et~al.}(2007)\citenamefont {Mazin},
  \citenamefont {Khomskii}, \citenamefont {Lengsdorf}, \citenamefont {Alonso},
  \citenamefont {Marshall}, \citenamefont {Ibberson}, \citenamefont
  {Podlesnyak}, \citenamefont {Mart\'{\i}nez-Lope},\ and\ \citenamefont
  {Abd-Elmeguid}}]{mazin2012}%
  \BibitemOpen
  \bibfield  {author} {\bibinfo {author} {\bibfnamefont {I.~I.}\ \bibnamefont
  {Mazin}}, \bibinfo {author} {\bibfnamefont {D.~I.}\ \bibnamefont {Khomskii}},
  \bibinfo {author} {\bibfnamefont {R.}~\bibnamefont {Lengsdorf}}, \bibinfo
  {author} {\bibfnamefont {J.~A.}\ \bibnamefont {Alonso}}, \bibinfo {author}
  {\bibfnamefont {W.~G.}\ \bibnamefont {Marshall}}, \bibinfo {author}
  {\bibfnamefont {R.~M.}\ \bibnamefont {Ibberson}}, \bibinfo {author}
  {\bibfnamefont {A.}~\bibnamefont {Podlesnyak}}, \bibinfo {author}
  {\bibfnamefont {M.~J.}\ \bibnamefont {Mart\'{\i}nez-Lope}}, \ and\ \bibinfo
  {author} {\bibfnamefont {M.~M.}\ \bibnamefont {Abd-Elmeguid}},\ }\href
  {\doibase 10.1103/PhysRevLett.98.176406} {\bibfield  {journal} {\bibinfo
  {journal} {Phys. Rev. Lett.}\ }\textbf {\bibinfo {volume} {98}},\ \bibinfo
  {pages} {176406} (\bibinfo {year} {2007})}\BibitemShut {NoStop}%
\bibitem [{\citenamefont {Johnston}\ \emph {et~al.}(2014)\citenamefont
  {Johnston}, \citenamefont {Mukherjee}, \citenamefont {Elfimov}, \citenamefont
  {Berciu},\ and\ \citenamefont {Sawatzky}}]{johnston2014}%
  \BibitemOpen
  \bibfield  {author} {\bibinfo {author} {\bibfnamefont {S.}~\bibnamefont
  {Johnston}}, \bibinfo {author} {\bibfnamefont {A.}~\bibnamefont {Mukherjee}},
  \bibinfo {author} {\bibfnamefont {I.}~\bibnamefont {Elfimov}}, \bibinfo
  {author} {\bibfnamefont {M.}~\bibnamefont {Berciu}}, \ and\ \bibinfo {author}
  {\bibfnamefont {G.~A.}\ \bibnamefont {Sawatzky}},\ }\href {\doibase
  10.1103/PhysRevLett.112.106404} {\bibfield  {journal} {\bibinfo  {journal}
  {Phys. Rev. Lett.}\ }\textbf {\bibinfo {volume} {112}},\ \bibinfo {pages}
  {106404} (\bibinfo {year} {2014})}\BibitemShut {NoStop}%
\bibitem [{\citenamefont {Subedi}\ \emph {et~al.}(2015)\citenamefont {Subedi},
  \citenamefont {Peil},\ and\ \citenamefont {Georges}}]{subedi2015}%
  \BibitemOpen
  \bibfield  {author} {\bibinfo {author} {\bibfnamefont {A.}~\bibnamefont
  {Subedi}}, \bibinfo {author} {\bibfnamefont {O.~E.}\ \bibnamefont {Peil}}, \
  and\ \bibinfo {author} {\bibfnamefont {A.}~\bibnamefont {Georges}},\ }\href
  {\doibase 10.1103/PhysRevB.91.075128} {\bibfield  {journal} {\bibinfo
  {journal} {Phys. Rev. B}\ }\textbf {\bibinfo {volume} {91}},\ \bibinfo
  {pages} {075128} (\bibinfo {year} {2015})}\BibitemShut {NoStop}%
\bibitem [{\citenamefont {Seth}\ \emph {et~al.}(2017)\citenamefont {Seth},
  \citenamefont {Peil}, \citenamefont {Pourovskii}, \citenamefont {Betzinger},
  \citenamefont {Friedrich}, \citenamefont {Parcollet}, \citenamefont
  {Biermann}, \citenamefont {Aryasetiawan},\ and\ \citenamefont
  {Georges}}]{seth2017}%
  \BibitemOpen
  \bibfield  {author} {\bibinfo {author} {\bibfnamefont {P.}~\bibnamefont
  {Seth}}, \bibinfo {author} {\bibfnamefont {O.~E.}\ \bibnamefont {Peil}},
  \bibinfo {author} {\bibfnamefont {L.}~\bibnamefont {Pourovskii}}, \bibinfo
  {author} {\bibfnamefont {M.}~\bibnamefont {Betzinger}}, \bibinfo {author}
  {\bibfnamefont {C.}~\bibnamefont {Friedrich}}, \bibinfo {author}
  {\bibfnamefont {O.}~\bibnamefont {Parcollet}}, \bibinfo {author}
  {\bibfnamefont {S.}~\bibnamefont {Biermann}}, \bibinfo {author}
  {\bibfnamefont {F.}~\bibnamefont {Aryasetiawan}}, \ and\ \bibinfo {author}
  {\bibfnamefont {A.}~\bibnamefont {Georges}},\ }\href {\doibase
  10.1103/PhysRevB.96.205139} {\bibfield  {journal} {\bibinfo  {journal} {Phys.
  Rev. B}\ }\textbf {\bibinfo {volume} {96}},\ \bibinfo {pages} {205139}
  (\bibinfo {year} {2017})}\BibitemShut {NoStop}%
\bibitem [{\citenamefont {Hirsch}(1985)}]{hirsch1985}%
  \BibitemOpen
  \bibfield  {author} {\bibinfo {author} {\bibfnamefont {J.~E.}\ \bibnamefont
  {Hirsch}},\ }\href {\doibase 10.1103/PhysRevLett.54.1317} {\bibfield
  {journal} {\bibinfo  {journal} {Phys. Rev. Lett.}\ }\textbf {\bibinfo
  {volume} {54}},\ \bibinfo {pages} {1317} (\bibinfo {year}
  {1985})}\BibitemShut {NoStop}%
\bibitem [{\citenamefont {van~der Marel}\ and\ \citenamefont
  {Sawatzky}(1985)}]{marel1985}%
  \BibitemOpen
  \bibfield  {author} {\bibinfo {author} {\bibfnamefont {D.}~\bibnamefont
  {van~der Marel}}\ and\ \bibinfo {author} {\bibfnamefont {G.~A.}\ \bibnamefont
  {Sawatzky}},\ }\href {\doibase 10.1016/0038-1098(85)90562-9} {\bibfield
  {journal} {\bibinfo  {journal} {Solid State Commun.}\ }\textbf {\bibinfo
  {volume} {55}},\ \bibinfo {pages} {937} (\bibinfo {year} {1985})}\BibitemShut
  {NoStop}%
\bibitem [{\citenamefont {Varma}(1988)}]{varma1988}%
  \BibitemOpen
  \bibfield  {author} {\bibinfo {author} {\bibfnamefont {C.~M.}\ \bibnamefont
  {Varma}},\ }\href {\doibase 10.1103/PhysRevLett.61.2713} {\bibfield
  {journal} {\bibinfo  {journal} {Phys. Rev. Lett.}\ }\textbf {\bibinfo
  {volume} {61}},\ \bibinfo {pages} {2713} (\bibinfo {year}
  {1988})}\BibitemShut {NoStop}%
\bibitem [{\citenamefont {Strand}(2014)}]{strand2014}%
  \BibitemOpen
  \bibfield  {author} {\bibinfo {author} {\bibfnamefont {H.~U.~R.}\
  \bibnamefont {Strand}},\ }\href {\doibase 10.1103/PhysRevB.90.155108}
  {\bibfield  {journal} {\bibinfo  {journal} {Phys. Rev. B}\ }\textbf {\bibinfo
  {volume} {90}},\ \bibinfo {pages} {155108} (\bibinfo {year}
  {2014})}\BibitemShut {NoStop}%
\bibitem [{\citenamefont {Fowlie}(2018)}]{fowlie2018}%
  \BibitemOpen
  \bibfield  {author} {\bibinfo {author} {\bibfnamefont {J.}~\bibnamefont
  {Fowlie}},\ }\emph {\bibinfo {title} {{Electronic and structural properties
  of LaNiO$_3$-based heterostructures}}},\ \href {\doibase
  10.13097/archive-ouverte/unige:120334} {Ph.D. thesis},\ \bibinfo  {school}
  {University of Geneva} (\bibinfo {year} {2018})\BibitemShut {NoStop}%
\bibitem [{\citenamefont {Li}\ \emph {et~al.}(2019)\citenamefont {Li},
  \citenamefont {Lee}, \citenamefont {Wang}, \citenamefont {Osada},
  \citenamefont {Crossley}, \citenamefont {Lee}, \citenamefont {Cui},
  \citenamefont {Hikita},\ and\ \citenamefont {Hwang}}]{li2019}%
  \BibitemOpen
  \bibfield  {author} {\bibinfo {author} {\bibfnamefont {D.}~\bibnamefont
  {Li}}, \bibinfo {author} {\bibfnamefont {K.}~\bibnamefont {Lee}}, \bibinfo
  {author} {\bibfnamefont {B.~Y.}\ \bibnamefont {Wang}}, \bibinfo {author}
  {\bibfnamefont {M.}~\bibnamefont {Osada}}, \bibinfo {author} {\bibfnamefont
  {S.}~\bibnamefont {Crossley}}, \bibinfo {author} {\bibfnamefont {H.~R.}\
  \bibnamefont {Lee}}, \bibinfo {author} {\bibfnamefont {Y.}~\bibnamefont
  {Cui}}, \bibinfo {author} {\bibfnamefont {Y.}~\bibnamefont {Hikita}}, \ and\
  \bibinfo {author} {\bibfnamefont {H.~Y.}\ \bibnamefont {Hwang}},\ }\href
  {\doibase 10.1038/s41586-019-1496-5} {\bibfield  {journal} {\bibinfo
  {journal} {Nature}\ }\textbf {\bibinfo {volume} {572}},\ \bibinfo {pages}
  {624} (\bibinfo {year} {2019})}\BibitemShut {NoStop}%
\bibitem [{\citenamefont {Catalano}\ \emph {et~al.}(2018)\citenamefont
  {Catalano}, \citenamefont {Gibert}, \citenamefont {Fowlie}, \citenamefont
  {{\'{I}}{\~{n}}iguez}, \citenamefont {Triscone},\ and\ \citenamefont
  {Kreisel}}]{catalano2018}%
  \BibitemOpen
  \bibfield  {author} {\bibinfo {author} {\bibfnamefont {S.}~\bibnamefont
  {Catalano}}, \bibinfo {author} {\bibfnamefont {M.}~\bibnamefont {Gibert}},
  \bibinfo {author} {\bibfnamefont {J.}~\bibnamefont {Fowlie}}, \bibinfo
  {author} {\bibfnamefont {J.}~\bibnamefont {{\'{I}}{\~{n}}iguez}}, \bibinfo
  {author} {\bibfnamefont {J.-M.}\ \bibnamefont {Triscone}}, \ and\ \bibinfo
  {author} {\bibfnamefont {J.}~\bibnamefont {Kreisel}},\ }\href {\doibase
  10.1088/1361-6633/aaa37a} {\bibfield  {journal} {\bibinfo  {journal} {Rep.
  Prog. Phys}\ }\textbf {\bibinfo {volume} {81}},\ \bibinfo {pages} {046501}
  (\bibinfo {year} {2018})}\BibitemShut {NoStop}%
\bibitem [{\citenamefont {Fowlie}\ \emph {et~al.}(2019)\citenamefont {Fowlie},
  \citenamefont {Lichtensteiger}, \citenamefont {Gibert}, \citenamefont
  {Meley}, \citenamefont {Willmott},\ and\ \citenamefont
  {Triscone}}]{fowlie2019}%
  \BibitemOpen
  \bibfield  {author} {\bibinfo {author} {\bibfnamefont {J.}~\bibnamefont
  {Fowlie}}, \bibinfo {author} {\bibfnamefont {C.}~\bibnamefont
  {Lichtensteiger}}, \bibinfo {author} {\bibfnamefont {M.}~\bibnamefont
  {Gibert}}, \bibinfo {author} {\bibfnamefont {H.}~\bibnamefont {Meley}},
  \bibinfo {author} {\bibfnamefont {P.}~\bibnamefont {Willmott}}, \ and\
  \bibinfo {author} {\bibfnamefont {J.-M.}\ \bibnamefont {Triscone}},\ }\href
  {\doibase 10.1021/acs.nanolett.9b01772} {\bibfield  {journal} {\bibinfo
  {journal} {Nano Lett.}\ }\textbf {\bibinfo {volume} {19}},\ \bibinfo {pages}
  {4188} (\bibinfo {year} {2019})}\BibitemShut {NoStop}%
\bibitem [{\citenamefont {Zhu}\ \emph {et~al.}(2013)\citenamefont {Zhu},
  \citenamefont {Komissinskiy}, \citenamefont {Radetinac}, \citenamefont
  {Vafaee}, \citenamefont {Wang},\ and\ \citenamefont {Alff}}]{zhu2013}%
  \BibitemOpen
  \bibfield  {author} {\bibinfo {author} {\bibfnamefont {M.}~\bibnamefont
  {Zhu}}, \bibinfo {author} {\bibfnamefont {P.}~\bibnamefont {Komissinskiy}},
  \bibinfo {author} {\bibfnamefont {A.}~\bibnamefont {Radetinac}}, \bibinfo
  {author} {\bibfnamefont {M.}~\bibnamefont {Vafaee}}, \bibinfo {author}
  {\bibfnamefont {Z.}~\bibnamefont {Wang}}, \ and\ \bibinfo {author}
  {\bibfnamefont {L.}~\bibnamefont {Alff}},\ }\href {\doibase
  10.1063/1.4823697} {\bibfield  {journal} {\bibinfo  {journal} {Applied Phys.
  Lett.}\ }\textbf {\bibinfo {volume} {103}},\ \bibinfo {pages} {141902}
  (\bibinfo {year} {2013})}\BibitemShut {NoStop}%
\bibitem [{\citenamefont {Guo}\ \emph {et~al.}(2018)\citenamefont {Guo},
  \citenamefont {Li}, \citenamefont {Hu}, \citenamefont {Kuo}, \citenamefont
  {Schmidt}, \citenamefont {Piovano}, \citenamefont {Pi}, \citenamefont
  {Sobolev}, \citenamefont {Khomskii}, \citenamefont {Tjeng},\ and\
  \citenamefont {Komarek}}]{guo2018}%
  \BibitemOpen
  \bibfield  {author} {\bibinfo {author} {\bibfnamefont {H.}~\bibnamefont
  {Guo}}, \bibinfo {author} {\bibfnamefont {L.}~\bibnamefont {Li},
  \bibfnamefont {W.~Z .and~Zhao}}, \bibinfo {author} {\bibfnamefont {C.~F.}\
  \bibnamefont {Hu}, \bibfnamefont {Z.and~Chang}}, \bibinfo {author}
  {\bibfnamefont {C.}~\bibnamefont {Kuo}}, \bibinfo {author} {\bibfnamefont
  {W.}~\bibnamefont {Schmidt}}, \bibinfo {author} {\bibfnamefont
  {A.}~\bibnamefont {Piovano}}, \bibinfo {author} {\bibfnamefont {T.-W.}\
  \bibnamefont {Pi}}, \bibinfo {author} {\bibfnamefont {O.}~\bibnamefont
  {Sobolev}}, \bibinfo {author} {\bibfnamefont {D.~I.}\ \bibnamefont
  {Khomskii}}, \bibinfo {author} {\bibfnamefont {L.~H.}\ \bibnamefont {Tjeng}},
  \ and\ \bibinfo {author} {\bibfnamefont {A.~C.}\ \bibnamefont {Komarek}},\
  }\href {\doibase 10.1038/s41467-017-02524-x} {\bibfield  {journal} {\bibinfo
  {journal} {Nat. Commun.}\ }\textbf {\bibinfo {volume} {9}},\ \bibinfo {pages}
  {43} (\bibinfo {year} {2018})}\BibitemShut {NoStop}%
\bibitem [{\citenamefont {Ouellette}\ \emph {et~al.}(2010)\citenamefont
  {Ouellette}, \citenamefont {Lee}, \citenamefont {Son}, \citenamefont
  {Stemmer}, \citenamefont {Balents}, \citenamefont {Millis},\ and\
  \citenamefont {Allen}}]{ouellette2010}%
  \BibitemOpen
  \bibfield  {author} {\bibinfo {author} {\bibfnamefont {D.~G.}\ \bibnamefont
  {Ouellette}}, \bibinfo {author} {\bibfnamefont {S.~B.}\ \bibnamefont {Lee}},
  \bibinfo {author} {\bibfnamefont {J.}~\bibnamefont {Son}}, \bibinfo {author}
  {\bibfnamefont {S.}~\bibnamefont {Stemmer}}, \bibinfo {author} {\bibfnamefont
  {L.}~\bibnamefont {Balents}}, \bibinfo {author} {\bibfnamefont {A.~J.}\
  \bibnamefont {Millis}}, \ and\ \bibinfo {author} {\bibfnamefont {S.~J.}\
  \bibnamefont {Allen}},\ }\href {\doibase 10.1103/PhysRevB.82.165112}
  {\bibfield  {journal} {\bibinfo  {journal} {Phys. Rev. B}\ }\textbf {\bibinfo
  {volume} {82}},\ \bibinfo {pages} {165112} (\bibinfo {year}
  {2010})}\BibitemShut {NoStop}%
\bibitem [{\citenamefont {Stewart}\ \emph {et~al.}(2011)\citenamefont
  {Stewart}, \citenamefont {Yee}, \citenamefont {Liu}, \citenamefont {Kareev},
  \citenamefont {Smith}, \citenamefont {Chapler}, \citenamefont {Varela},
  \citenamefont {Ryan}, \citenamefont {Haule}, \citenamefont {Chakhalian},\
  and\ \citenamefont {Basov}}]{stewart2011}%
  \BibitemOpen
  \bibfield  {author} {\bibinfo {author} {\bibfnamefont {M.~K.}\ \bibnamefont
  {Stewart}}, \bibinfo {author} {\bibfnamefont {C.-H.}\ \bibnamefont {Yee}},
  \bibinfo {author} {\bibfnamefont {J.}~\bibnamefont {Liu}}, \bibinfo {author}
  {\bibfnamefont {M.}~\bibnamefont {Kareev}}, \bibinfo {author} {\bibfnamefont
  {R.~K.}\ \bibnamefont {Smith}}, \bibinfo {author} {\bibfnamefont {B.~C.}\
  \bibnamefont {Chapler}}, \bibinfo {author} {\bibfnamefont {M.}~\bibnamefont
  {Varela}}, \bibinfo {author} {\bibfnamefont {P.~J.}\ \bibnamefont {Ryan}},
  \bibinfo {author} {\bibfnamefont {K.}~\bibnamefont {Haule}}, \bibinfo
  {author} {\bibfnamefont {J.}~\bibnamefont {Chakhalian}}, \ and\ \bibinfo
  {author} {\bibfnamefont {D.~N.}\ \bibnamefont {Basov}},\ }\href {\doibase
  10.1103/PhysRevB.83.075125} {\bibfield  {journal} {\bibinfo  {journal} {Phys.
  Rev. B}\ }\textbf {\bibinfo {volume} {83}},\ \bibinfo {pages} {075125}
  (\bibinfo {year} {2011})}\BibitemShut {NoStop}%
\bibitem [{\citenamefont {Yoo}\ \emph {et~al.}(2015)\citenamefont {Yoo},
  \citenamefont {Hyun}, \citenamefont {Moreschini}, \citenamefont {Kim},
  \citenamefont {Chang}, \citenamefont {Sohn}, \citenamefont {Jeong},
  \citenamefont {Sinn}, \citenamefont {Kim}, \citenamefont {Bostwick},
  \citenamefont {Rotenberg}, \citenamefont {Shim},\ and\ \citenamefont
  {Noh}}]{Yoo2015}%
  \BibitemOpen
  \bibfield  {author} {\bibinfo {author} {\bibfnamefont {H.~K.}\ \bibnamefont
  {Yoo}}, \bibinfo {author} {\bibfnamefont {S.~I.}\ \bibnamefont {Hyun}},
  \bibinfo {author} {\bibfnamefont {L.}~\bibnamefont {Moreschini}}, \bibinfo
  {author} {\bibfnamefont {H.-D.}\ \bibnamefont {Kim}}, \bibinfo {author}
  {\bibfnamefont {Y.~J.}\ \bibnamefont {Chang}}, \bibinfo {author}
  {\bibfnamefont {C.~H.}\ \bibnamefont {Sohn}}, \bibinfo {author}
  {\bibfnamefont {D.~W.}\ \bibnamefont {Jeong}}, \bibinfo {author}
  {\bibfnamefont {S.}~\bibnamefont {Sinn}}, \bibinfo {author} {\bibfnamefont
  {Y.~S.}\ \bibnamefont {Kim}}, \bibinfo {author} {\bibfnamefont
  {A.}~\bibnamefont {Bostwick}}, \bibinfo {author} {\bibfnamefont
  {E.}~\bibnamefont {Rotenberg}}, \bibinfo {author} {\bibfnamefont {J.~H.}\
  \bibnamefont {Shim}}, \ and\ \bibinfo {author} {\bibfnamefont {T.~W.}\
  \bibnamefont {Noh}},\ }\href {\doibase 10.1038/srep08746} {\bibfield
  {journal} {\bibinfo  {journal} {Sci. Rep.}\ }\textbf {\bibinfo {volume}
  {5}},\ \bibinfo {pages} {8746} (\bibinfo {year} {2015})}\BibitemShut
  {NoStop}%
\bibitem [{\citenamefont {Kuzmenko}(2005)}]{kuzmenko2005}%
  \BibitemOpen
  \bibfield  {author} {\bibinfo {author} {\bibfnamefont {A.}~\bibnamefont
  {Kuzmenko}},\ }\href {\doibase 10.1063/1.1979470} {\bibfield  {journal}
  {\bibinfo  {journal} {Rev. Sci. Instrum.}\ }\textbf {\bibinfo {volume}
  {76}},\ \bibinfo {pages} {083108} (\bibinfo {year} {2005})}\BibitemShut
  {NoStop}%
\bibitem [{\citenamefont {Nakajima}\ \emph {et~al.}(2010)\citenamefont
  {Nakajima}, \citenamefont {Ishida}, \citenamefont {Kihou}, \citenamefont
  {Tomioka}, \citenamefont {Ito}, \citenamefont {Yoshida}, \citenamefont {Lee},
  \citenamefont {Kito}, \citenamefont {Iyo}, \citenamefont {Eisaki},
  \citenamefont {M.~Kojima},\ and\ \citenamefont {Uchida}}]{nakajima2010}%
  \BibitemOpen
  \bibfield  {author} {\bibinfo {author} {\bibfnamefont {M.}~\bibnamefont
  {Nakajima}}, \bibinfo {author} {\bibfnamefont {S.}~\bibnamefont {Ishida}},
  \bibinfo {author} {\bibfnamefont {K.}~\bibnamefont {Kihou}}, \bibinfo
  {author} {\bibfnamefont {Y.}~\bibnamefont {Tomioka}}, \bibinfo {author}
  {\bibfnamefont {T.}~\bibnamefont {Ito}}, \bibinfo {author} {\bibfnamefont
  {Y.}~\bibnamefont {Yoshida}}, \bibinfo {author} {\bibfnamefont {C.-H.}\
  \bibnamefont {Lee}}, \bibinfo {author} {\bibfnamefont {H.}~\bibnamefont
  {Kito}}, \bibinfo {author} {\bibfnamefont {A.}~\bibnamefont {Iyo}}, \bibinfo
  {author} {\bibfnamefont {H.}~\bibnamefont {Eisaki}}, \bibinfo {author}
  {\bibfnamefont {K.}~\bibnamefont {M.~Kojima}}, \ and\ \bibinfo {author}
  {\bibfnamefont {S.-I.}\ \bibnamefont {Uchida}},\ }\href {\doibase
  10.1103/PhysRevB.81.104528} {\bibfield  {journal} {\bibinfo  {journal} {Phys.
  Rev. B}\ }\textbf {\bibinfo {volume} {81}},\ \bibinfo {pages} {104528}
  (\bibinfo {year} {2010})}\BibitemShut {NoStop}%
\bibitem [{\citenamefont {Cheng}\ \emph {et~al.}(2012)\citenamefont {Cheng},
  \citenamefont {Hu}, \citenamefont {Chen}, \citenamefont {Xu}, \citenamefont
  {Zheng}, \citenamefont {Luo},\ and\ \citenamefont {Wang}}]{cheng2012}%
  \BibitemOpen
  \bibfield  {author} {\bibinfo {author} {\bibfnamefont {B.}~\bibnamefont
  {Cheng}}, \bibinfo {author} {\bibfnamefont {B.~F.}\ \bibnamefont {Hu}},
  \bibinfo {author} {\bibfnamefont {R.~Y.}\ \bibnamefont {Chen}}, \bibinfo
  {author} {\bibfnamefont {G.}~\bibnamefont {Xu}}, \bibinfo {author}
  {\bibfnamefont {P.}~\bibnamefont {Zheng}}, \bibinfo {author} {\bibfnamefont
  {J.~L.}\ \bibnamefont {Luo}}, \ and\ \bibinfo {author} {\bibfnamefont
  {N.~L.}\ \bibnamefont {Wang}},\ }\href {\doibase 10.1103/PhysRevB.86.134503}
  {\bibfield  {journal} {\bibinfo  {journal} {Phys. Rev. B}\ }\textbf {\bibinfo
  {volume} {86}},\ \bibinfo {pages} {134503} (\bibinfo {year}
  {2012})}\BibitemShut {NoStop}%
\bibitem [{\citenamefont {Ruppen}\ \emph {et~al.}(2015)\citenamefont {Ruppen},
  \citenamefont {Teyssier}, \citenamefont {Peil}, \citenamefont {Catalano},
  \citenamefont {Gibert}, \citenamefont {Mravlje}, \citenamefont {Triscone},
  \citenamefont {Georges},\ and\ \citenamefont {van~der Marel}}]{ruppen2015}%
  \BibitemOpen
  \bibfield  {author} {\bibinfo {author} {\bibfnamefont {J.}~\bibnamefont
  {Ruppen}}, \bibinfo {author} {\bibfnamefont {J.}~\bibnamefont {Teyssier}},
  \bibinfo {author} {\bibfnamefont {O.~E.}\ \bibnamefont {Peil}}, \bibinfo
  {author} {\bibfnamefont {S.}~\bibnamefont {Catalano}}, \bibinfo {author}
  {\bibfnamefont {M.}~\bibnamefont {Gibert}}, \bibinfo {author} {\bibfnamefont
  {J.}~\bibnamefont {Mravlje}}, \bibinfo {author} {\bibfnamefont {J.-M.}\
  \bibnamefont {Triscone}}, \bibinfo {author} {\bibfnamefont {A.}~\bibnamefont
  {Georges}}, \ and\ \bibinfo {author} {\bibfnamefont {D.}~\bibnamefont
  {van~der Marel}},\ }\href {\doibase 10.1103/PhysRevB.92.155145} {\bibfield
  {journal} {\bibinfo  {journal} {Phys. Rev. B}\ }\textbf {\bibinfo {volume}
  {92}},\ \bibinfo {pages} {155145} (\bibinfo {year} {2015})}\BibitemShut
  {NoStop}%
\bibitem [{\citenamefont {Berthod}\ \emph {et~al.}(2013)\citenamefont
  {Berthod}, \citenamefont {Mravlje}, \citenamefont {Deng}, \citenamefont
  {Zitko}, \citenamefont {van~der Marel},\ and\ \citenamefont
  {Georges}}]{berthod2013}%
  \BibitemOpen
  \bibfield  {author} {\bibinfo {author} {\bibfnamefont {C.}~\bibnamefont
  {Berthod}}, \bibinfo {author} {\bibfnamefont {J.}~\bibnamefont {Mravlje}},
  \bibinfo {author} {\bibfnamefont {X.}~\bibnamefont {Deng}}, \bibinfo {author}
  {\bibfnamefont {R.}~\bibnamefont {Zitko}}, \bibinfo {author} {\bibfnamefont
  {D.}~\bibnamefont {van~der Marel}}, \ and\ \bibinfo {author} {\bibfnamefont
  {A.}~\bibnamefont {Georges}},\ }\href {\doibase 10.1103/PhysRevB.87.115109}
  {\bibfield  {journal} {\bibinfo  {journal} {Phys. Rev. B}\ }\textbf {\bibinfo
  {volume} {87}},\ \bibinfo {pages} {115109} (\bibinfo {year}
  {2013})}\BibitemShut {NoStop}%
\bibitem [{\citenamefont {G\"otze}\ and\ \citenamefont
  {W\"olfle}(1972)}]{gotze1972}%
  \BibitemOpen
  \bibfield  {author} {\bibinfo {author} {\bibfnamefont {W.}~\bibnamefont
  {G\"otze}}\ and\ \bibinfo {author} {\bibfnamefont {P.}~\bibnamefont
  {W\"olfle}},\ }\href {\doibase 10.1103/PhysRevB.6.1226} {\bibfield  {journal}
  {\bibinfo  {journal} {Phys. Rev. B}\ }\textbf {\bibinfo {volume} {6}},\
  \bibinfo {pages} {1226} (\bibinfo {year} {1972})}\BibitemShut {NoStop}%
\bibitem [{\citenamefont {Sreedhar}\ \emph {et~al.}(1992)\citenamefont
  {Sreedhar}, \citenamefont {Honig}, \citenamefont {Darwin}, \citenamefont
  {McElfresh}, \citenamefont {Shand}, \citenamefont {Xu}, \citenamefont
  {Crooker},\ and\ \citenamefont {Spalek}}]{Sreedhar1992}%
  \BibitemOpen
  \bibfield  {author} {\bibinfo {author} {\bibfnamefont {K.}~\bibnamefont
  {Sreedhar}}, \bibinfo {author} {\bibfnamefont {J.~M.}\ \bibnamefont {Honig}},
  \bibinfo {author} {\bibfnamefont {M.}~\bibnamefont {Darwin}}, \bibinfo
  {author} {\bibfnamefont {M.}~\bibnamefont {McElfresh}}, \bibinfo {author}
  {\bibfnamefont {P.~M.}\ \bibnamefont {Shand}}, \bibinfo {author}
  {\bibfnamefont {J.}~\bibnamefont {Xu}}, \bibinfo {author} {\bibfnamefont
  {B.~C.}\ \bibnamefont {Crooker}}, \ and\ \bibinfo {author} {\bibfnamefont
  {J.}~\bibnamefont {Spalek}},\ }\href {\doibase 10.1103/PhysRevB.46.6382}
  {\bibfield  {journal} {\bibinfo  {journal} {Phys. Rev. B}\ }\textbf {\bibinfo
  {volume} {46}},\ \bibinfo {pages} {6382} (\bibinfo {year}
  {1992})}\BibitemShut {NoStop}%
\bibitem [{\citenamefont {Son}\ \emph {et~al.}(2010)\citenamefont {Son},
  \citenamefont {Moetakef}, \citenamefont {LeBeau}, \citenamefont {Ouellette},
  \citenamefont {Balents}, \citenamefont {Allen},\ and\ \citenamefont
  {Stemmer}}]{son2010}%
  \BibitemOpen
  \bibfield  {author} {\bibinfo {author} {\bibfnamefont {J.}~\bibnamefont
  {Son}}, \bibinfo {author} {\bibfnamefont {P.}~\bibnamefont {Moetakef}},
  \bibinfo {author} {\bibfnamefont {J.~M.}\ \bibnamefont {LeBeau}}, \bibinfo
  {author} {\bibfnamefont {D.}~\bibnamefont {Ouellette}}, \bibinfo {author}
  {\bibfnamefont {L.}~\bibnamefont {Balents}}, \bibinfo {author} {\bibfnamefont
  {S.~J.}\ \bibnamefont {Allen}}, \ and\ \bibinfo {author} {\bibfnamefont
  {S.}~\bibnamefont {Stemmer}},\ }\href {\doibase 10.1063/1.3309713} {\bibfield
   {journal} {\bibinfo  {journal} {Applied Phys. Lett.}\ }\textbf {\bibinfo
  {volume} {96}},\ \bibinfo {pages} {062114} (\bibinfo {year}
  {2010})}\BibitemShut {NoStop}%
\bibitem [{\citenamefont {Scherwitzl}\ \emph {et~al.}(2011)\citenamefont
  {Scherwitzl}, \citenamefont {Gariglio}, \citenamefont {Gabay}, \citenamefont
  {Zubko}, \citenamefont {Gibert},\ and\ \citenamefont
  {Triscone}}]{Scherwitzl2011}%
  \BibitemOpen
  \bibfield  {author} {\bibinfo {author} {\bibfnamefont {R.}~\bibnamefont
  {Scherwitzl}}, \bibinfo {author} {\bibfnamefont {S.}~\bibnamefont
  {Gariglio}}, \bibinfo {author} {\bibfnamefont {M.}~\bibnamefont {Gabay}},
  \bibinfo {author} {\bibfnamefont {P.}~\bibnamefont {Zubko}}, \bibinfo
  {author} {\bibfnamefont {M.}~\bibnamefont {Gibert}}, \ and\ \bibinfo {author}
  {\bibfnamefont {J.-M.}\ \bibnamefont {Triscone}},\ }\href {\doibase
  10.1103/PhysRevLett.106.246403} {\bibfield  {journal} {\bibinfo  {journal}
  {Phys. Rev. Lett.}\ }\textbf {\bibinfo {volume} {106}},\ \bibinfo {pages}
  {246403} (\bibinfo {year} {2011})}\BibitemShut {NoStop}%
\bibitem [{\citenamefont {Scherwitzl}(2012)}]{scherwitzl2012}%
  \BibitemOpen
  \bibfield  {author} {\bibinfo {author} {\bibfnamefont {R.}~\bibnamefont
  {Scherwitzl}},\ }\emph {\bibinfo {title} {{Metal-insulator transitions in
  nickelate heterostructures}}},\ \href {\doibase
  10.13097/archive-ouverte/unige:21740} {Ph.D. thesis},\ \bibinfo  {school}
  {University of Geneva} (\bibinfo {year} {2012})\BibitemShut {NoStop}%
\bibitem [{\citenamefont {Liu}\ \emph {et~al.}(2020)\citenamefont {Liu},
  \citenamefont {Humbert}, \citenamefont {Bretz-Sullivan}, \citenamefont
  {Wang}, \citenamefont {Hong}, \citenamefont {Wrobel}, \citenamefont {Zhang},
  \citenamefont {Hoffman}, \citenamefont {Pearson}, \citenamefont {Jiang},
  \citenamefont {Chang}, \citenamefont {Suslov}, \citenamefont {Mason},
  \citenamefont {Norman},\ and\ \citenamefont {Bhattacharya}}]{Liu2020}%
  \BibitemOpen
  \bibfield  {author} {\bibinfo {author} {\bibfnamefont {C.}~\bibnamefont
  {Liu}}, \bibinfo {author} {\bibfnamefont {V.~F.~C.}\ \bibnamefont {Humbert}},
  \bibinfo {author} {\bibfnamefont {T.~M.}\ \bibnamefont {Bretz-Sullivan}},
  \bibinfo {author} {\bibfnamefont {G.}~\bibnamefont {Wang}}, \bibinfo {author}
  {\bibfnamefont {D.}~\bibnamefont {Hong}}, \bibinfo {author} {\bibfnamefont
  {F.}~\bibnamefont {Wrobel}}, \bibinfo {author} {\bibfnamefont
  {J.}~\bibnamefont {Zhang}}, \bibinfo {author} {\bibfnamefont {J.~D.}\
  \bibnamefont {Hoffman}}, \bibinfo {author} {\bibfnamefont {J.~E.}\
  \bibnamefont {Pearson}}, \bibinfo {author} {\bibfnamefont {J.~S.}\
  \bibnamefont {Jiang}}, \bibinfo {author} {\bibfnamefont {C.}~\bibnamefont
  {Chang}}, \bibinfo {author} {\bibfnamefont {A.}~\bibnamefont {Suslov}},
  \bibinfo {author} {\bibfnamefont {N.}~\bibnamefont {Mason}}, \bibinfo
  {author} {\bibfnamefont {M.~R.}\ \bibnamefont {Norman}}, \ and\ \bibinfo
  {author} {\bibfnamefont {A.}~\bibnamefont {Bhattacharya}},\ }\href {\doibase
  10.1038/s41467-020-15143-w} {\bibfield  {journal} {\bibinfo  {journal}
  {Nature Commun.}\ }\textbf {\bibinfo {volume} {11}},\ \bibinfo {pages} {1402}
  (\bibinfo {year} {2020})}\BibitemShut {NoStop}%
\bibitem [{Note1()}]{Note1}%
  \BibitemOpen
  \bibinfo {note} {Since the measured optical response depends on the film
  thickness, one may suspect that this discrepancy could result from assuming
  the wrong value for the film thickness, or from the presence of a dead
  ({\protect \em i.e.} insulating) layer. However, the thicknesses of these
  films are accurately known from X-ray diffraction. Since even for the
  insulating phase of RNiO$_3$ the spectral weight in the visible part of the
  spectrum is not so different from the metallic phase~\cite {ruppen2015}, the
  assumption of a dead layer cannot explain the low value of N$_{eff}$(2eV) for
  the tensile samples. Moreover, transport experiments on films on LaAlO$_3$ as
  a function of film thickness have demonstrated that the films switch as a
  whole from metal to insulator at a critical thickness~\cite
  {fowlie2017}.}\BibitemShut {Stop}%
\bibitem [{\citenamefont {King}\ \emph {et~al.}(2014)\citenamefont {King},
  \citenamefont {Wei}, \citenamefont {Nie}, \citenamefont {Uchida},
  \citenamefont {Adamo}, \citenamefont {Zhu}, \citenamefont {He}, \citenamefont
  {Bozovic}, \citenamefont {D.G.},\ and\ \citenamefont {Shen}}]{king2014}%
  \BibitemOpen
  \bibfield  {author} {\bibinfo {author} {\bibfnamefont {P.~D.~C.}\
  \bibnamefont {King}}, \bibinfo {author} {\bibfnamefont {H.}~\bibnamefont
  {Wei}}, \bibinfo {author} {\bibfnamefont {Y.~F.}\ \bibnamefont {Nie}},
  \bibinfo {author} {\bibfnamefont {M.}~\bibnamefont {Uchida}}, \bibinfo
  {author} {\bibfnamefont {C.}~\bibnamefont {Adamo}}, \bibinfo {author}
  {\bibfnamefont {S.}~\bibnamefont {Zhu}}, \bibinfo {author} {\bibfnamefont
  {X.}~\bibnamefont {He}}, \bibinfo {author} {\bibfnamefont {I.}~\bibnamefont
  {Bozovic}}, \bibinfo {author} {\bibfnamefont {S.}~\bibnamefont {D.G.}}, \
  and\ \bibinfo {author} {\bibfnamefont {K.}~\bibnamefont {Shen}},\ }\href
  {\doibase 10.1038/nnano.2014.59} {\bibfield  {journal} {\bibinfo  {journal}
  {Nat. Nanotechnol.}\ }\textbf {\bibinfo {volume} {9}},\ \bibinfo {pages}
  {443} (\bibinfo {year} {2014})}\BibitemShut {NoStop}%
\bibitem [{\citenamefont {Deng}\ \emph {et~al.}(2012)\citenamefont {Deng},
  \citenamefont {Ferrero}, \citenamefont {Mravlje}, \citenamefont {Aichhorn},\
  and\ \citenamefont {Georges}}]{deng_prb_2012}%
  \BibitemOpen
  \bibfield  {author} {\bibinfo {author} {\bibfnamefont {X.}~\bibnamefont
  {Deng}}, \bibinfo {author} {\bibfnamefont {M.}~\bibnamefont {Ferrero}},
  \bibinfo {author} {\bibfnamefont {J.}~\bibnamefont {Mravlje}}, \bibinfo
  {author} {\bibfnamefont {M.}~\bibnamefont {Aichhorn}}, \ and\ \bibinfo
  {author} {\bibfnamefont {A.}~\bibnamefont {Georges}},\ }\href {\doibase
  10.1103/PhysRevB.85.125137} {\bibfield  {journal} {\bibinfo  {journal} {Phys.
  Rev. B}\ }\textbf {\bibinfo {volume} {85}},\ \bibinfo {pages} {125137}
  (\bibinfo {year} {2012})}\BibitemShut {NoStop}%
\bibitem [{\citenamefont {Peil}\ \emph {et~al.}(2014)\citenamefont {Peil},
  \citenamefont {Ferrero},\ and\ \citenamefont {Georges}}]{peil2014}%
  \BibitemOpen
  \bibfield  {author} {\bibinfo {author} {\bibfnamefont {O.~E.}\ \bibnamefont
  {Peil}}, \bibinfo {author} {\bibfnamefont {M.}~\bibnamefont {Ferrero}}, \
  and\ \bibinfo {author} {\bibfnamefont {A.}~\bibnamefont {Georges}},\ }\href
  {\doibase 10.1103/PhysRevB.90.045128} {\bibfield  {journal} {\bibinfo
  {journal} {Phys. Rev. B}\ }\textbf {\bibinfo {volume} {90}},\ \bibinfo
  {pages} {045128} (\bibinfo {year} {2014})}\BibitemShut {NoStop}%
\bibitem [{\citenamefont {Nowadnick}\ \emph {et~al.}(2015)\citenamefont
  {Nowadnick}, \citenamefont {Ruf}, \citenamefont {Park}, \citenamefont {King},
  \citenamefont {Schlom}, \citenamefont {Shen},\ and\ \citenamefont
  {Millis}}]{nowadnick2015}%
  \BibitemOpen
  \bibfield  {author} {\bibinfo {author} {\bibfnamefont {E.~A.}\ \bibnamefont
  {Nowadnick}}, \bibinfo {author} {\bibfnamefont {J.~P.}\ \bibnamefont {Ruf}},
  \bibinfo {author} {\bibfnamefont {H.}~\bibnamefont {Park}}, \bibinfo {author}
  {\bibfnamefont {P.~D.~C.}\ \bibnamefont {King}}, \bibinfo {author}
  {\bibfnamefont {D.~G.}\ \bibnamefont {Schlom}}, \bibinfo {author}
  {\bibfnamefont {K.~M.}\ \bibnamefont {Shen}}, \ and\ \bibinfo {author}
  {\bibfnamefont {A.~J.}\ \bibnamefont {Millis}},\ }\href {\doibase
  10.1103/PhysRevB.92.245109} {\bibfield  {journal} {\bibinfo  {journal} {Phys.
  Rev. B}\ }\textbf {\bibinfo {volume} {92}},\ \bibinfo {pages} {245109}
  (\bibinfo {year} {2015})}\BibitemShut {NoStop}%
\bibitem [{\citenamefont {Basov}\ \emph {et~al.}(2011)\citenamefont {Basov},
  \citenamefont {Averitt}, \citenamefont {van~der Marel}, \citenamefont
  {Dressel},\ and\ \citenamefont {Haule}}]{basov2011}%
  \BibitemOpen
  \bibfield  {author} {\bibinfo {author} {\bibfnamefont {D.~N.}\ \bibnamefont
  {Basov}}, \bibinfo {author} {\bibfnamefont {R.~D.}\ \bibnamefont {Averitt}},
  \bibinfo {author} {\bibfnamefont {D.}~\bibnamefont {van~der Marel}}, \bibinfo
  {author} {\bibfnamefont {M.}~\bibnamefont {Dressel}}, \ and\ \bibinfo
  {author} {\bibfnamefont {K.}~\bibnamefont {Haule}},\ }\href {\doibase
  10.1103/RevModPhys.83.471} {\bibfield  {journal} {\bibinfo  {journal} {Rev.
  Mod. Phys.}\ }\textbf {\bibinfo {volume} {83}},\ \bibinfo {pages} {471}
  (\bibinfo {year} {2011})}\BibitemShut {NoStop}%
\bibitem [{\citenamefont {May}\ \emph {et~al.}(2010)\citenamefont {May},
  \citenamefont {Kim}, \citenamefont {Rondinelli}, \citenamefont {Karapetrova},
  \citenamefont {Spaldin}, \citenamefont {Bhattacharya},\ and\ \citenamefont
  {Ryan}}]{may2010}%
  \BibitemOpen
  \bibfield  {author} {\bibinfo {author} {\bibfnamefont {S.~J.}\ \bibnamefont
  {May}}, \bibinfo {author} {\bibfnamefont {J.-W.}\ \bibnamefont {Kim}},
  \bibinfo {author} {\bibfnamefont {J.~M.}\ \bibnamefont {Rondinelli}},
  \bibinfo {author} {\bibfnamefont {E.}~\bibnamefont {Karapetrova}}, \bibinfo
  {author} {\bibfnamefont {N.~A.}\ \bibnamefont {Spaldin}}, \bibinfo {author}
  {\bibfnamefont {A.}~\bibnamefont {Bhattacharya}}, \ and\ \bibinfo {author}
  {\bibfnamefont {P.~J.}\ \bibnamefont {Ryan}},\ }\href {\doibase
  10.1103/PhysRevB.82.014110} {\bibfield  {journal} {\bibinfo  {journal} {Phys.
  Rev. B}\ }\textbf {\bibinfo {volume} {82}},\ \bibinfo {pages} {014110}
  (\bibinfo {year} {2010})}\BibitemShut {NoStop}%
\bibitem [{\citenamefont {Marzari}\ and\ \citenamefont
  {Vanderbilt}(1997)}]{MLWF1}%
  \BibitemOpen
  \bibfield  {author} {\bibinfo {author} {\bibfnamefont {N.}~\bibnamefont
  {Marzari}}\ and\ \bibinfo {author} {\bibfnamefont {D.}~\bibnamefont
  {Vanderbilt}},\ }\href {\doibase 10.1103/PhysRevB.56.12847} {\bibfield
  {journal} {\bibinfo  {journal} {Phys. Rev. B}\ }\textbf {\bibinfo {volume}
  {56}},\ \bibinfo {pages} {12847} (\bibinfo {year} {1997})}\BibitemShut
  {NoStop}%
\bibitem [{\citenamefont {Souza}\ \emph {et~al.}(2001)\citenamefont {Souza},
  \citenamefont {Marzari},\ and\ \citenamefont {Vanderbilt}}]{MLWF2}%
  \BibitemOpen
  \bibfield  {author} {\bibinfo {author} {\bibfnamefont {I.}~\bibnamefont
  {Souza}}, \bibinfo {author} {\bibfnamefont {N.}~\bibnamefont {Marzari}}, \
  and\ \bibinfo {author} {\bibfnamefont {D.}~\bibnamefont {Vanderbilt}},\
  }\href {\doibase 10.1103/PhysRevB.65.035109} {\bibfield  {journal} {\bibinfo
  {journal} {Phys. Rev. B}\ }\textbf {\bibinfo {volume} {65}},\ \bibinfo
  {pages} {035109} (\bibinfo {year} {2001})}\BibitemShut {NoStop}%
\bibitem [{\citenamefont {Pizzi}\ \emph {et~al.}(2019)\citenamefont {Pizzi},
  \citenamefont {Vitale}, \citenamefont {Arita}, \citenamefont {Bluegel},
  \citenamefont {Freimuth}, \citenamefont {Geranton}, \citenamefont
  {Gibertini}, \citenamefont {Gresch}, \citenamefont {Johnson}, \citenamefont
  {Koretsune}, \citenamefont {Ibanez}, \citenamefont {Lee}, \citenamefont
  {Lihm}, \citenamefont {Marchand}, \citenamefont {Marrazzo}, \citenamefont
  {Mokrousov}, \citenamefont {Mustafa}, \citenamefont {Nohara}, \citenamefont
  {Nomura}, \citenamefont {Paulatto}, \citenamefont {Ponce}, \citenamefont
  {Ponweiser}, \citenamefont {Qiao}, \citenamefont {Thole}, \citenamefont
  {Tsirkin}, \citenamefont {Wierzbowska}, \citenamefont {Marzari},
  \citenamefont {Vanderbilt}, \citenamefont {Souza}, \citenamefont {Mostofi},\
  and\ \citenamefont {Yates}}]{wannier90}%
  \BibitemOpen
  \bibfield  {author} {\bibinfo {author} {\bibfnamefont {G.}~\bibnamefont
  {Pizzi}}, \bibinfo {author} {\bibfnamefont {V.}~\bibnamefont {Vitale}},
  \bibinfo {author} {\bibfnamefont {R.}~\bibnamefont {Arita}}, \bibinfo
  {author} {\bibfnamefont {S.}~\bibnamefont {Bluegel}}, \bibinfo {author}
  {\bibfnamefont {F.}~\bibnamefont {Freimuth}}, \bibinfo {author}
  {\bibfnamefont {G.}~\bibnamefont {Geranton}}, \bibinfo {author}
  {\bibfnamefont {M.}~\bibnamefont {Gibertini}}, \bibinfo {author}
  {\bibfnamefont {D.}~\bibnamefont {Gresch}}, \bibinfo {author} {\bibfnamefont
  {C.}~\bibnamefont {Johnson}}, \bibinfo {author} {\bibfnamefont
  {T.}~\bibnamefont {Koretsune}}, \bibinfo {author} {\bibfnamefont
  {J.}~\bibnamefont {Ibanez}}, \bibinfo {author} {\bibfnamefont
  {H.}~\bibnamefont {Lee}}, \bibinfo {author} {\bibfnamefont {J.-M.}\
  \bibnamefont {Lihm}}, \bibinfo {author} {\bibfnamefont {D.}~\bibnamefont
  {Marchand}}, \bibinfo {author} {\bibfnamefont {A.}~\bibnamefont {Marrazzo}},
  \bibinfo {author} {\bibfnamefont {Y.}~\bibnamefont {Mokrousov}}, \bibinfo
  {author} {\bibfnamefont {J.~I.}\ \bibnamefont {Mustafa}}, \bibinfo {author}
  {\bibfnamefont {Y.}~\bibnamefont {Nohara}}, \bibinfo {author} {\bibfnamefont
  {Y.}~\bibnamefont {Nomura}}, \bibinfo {author} {\bibfnamefont
  {L.}~\bibnamefont {Paulatto}}, \bibinfo {author} {\bibfnamefont
  {S.}~\bibnamefont {Ponce}}, \bibinfo {author} {\bibfnamefont
  {T.}~\bibnamefont {Ponweiser}}, \bibinfo {author} {\bibfnamefont
  {J.}~\bibnamefont {Qiao}}, \bibinfo {author} {\bibfnamefont {F.}~\bibnamefont
  {Thole}}, \bibinfo {author} {\bibfnamefont {S.~S.}\ \bibnamefont {Tsirkin}},
  \bibinfo {author} {\bibfnamefont {M.}~\bibnamefont {Wierzbowska}}, \bibinfo
  {author} {\bibfnamefont {N.}~\bibnamefont {Marzari}}, \bibinfo {author}
  {\bibfnamefont {D.}~\bibnamefont {Vanderbilt}}, \bibinfo {author}
  {\bibfnamefont {I.}~\bibnamefont {Souza}}, \bibinfo {author} {\bibfnamefont
  {A.~A.}\ \bibnamefont {Mostofi}}, \ and\ \bibinfo {author} {\bibfnamefont
  {J.~R.}\ \bibnamefont {Yates}},\ }\href {\doibase 10.1088/1361-648X/ab51ff}
  {\bibfield  {journal} {\bibinfo  {journal} {J. Phys. Condens. Matter}\
  }\textbf {\bibinfo {volume} {32}},\ \bibinfo {pages} {165902} (\bibinfo
  {year} {2019})}\BibitemShut {NoStop}%
\bibitem [{Bin()}]{Binder}%
  \BibitemOpen
  \href@noop {} {\enquote {\bibinfo {title} {{Online Interactive Fermi Surface
  Applet}},}\ }\bibinfo {howpublished}
  {\url{https://lnostrain.page.link/applet} Note: It will take a few minutes
  until the application is loaded and built if it has not been used by anyone
  recently. Once loaded, click on RUN\_ME.ipynb and follow the instructions
  therein.}\BibitemShut {Stop}%
\bibitem [{\citenamefont {Lichtensteiger}(2018)}]{Lichtensteiger2018}%
  \BibitemOpen
  \bibfield  {author} {\bibinfo {author} {\bibfnamefont {C.}~\bibnamefont
  {Lichtensteiger}},\ }\href {\doibase 10.1107/S1600576718012840} {\bibfield
  {journal} {\bibinfo  {journal} {Journal of Applied Crystallography}\ }\textbf
  {\bibinfo {volume} {51}},\ \bibinfo {pages} {1745} (\bibinfo {year}
  {2018})}\BibitemShut {NoStop}%
\bibitem [{\citenamefont {Perdew}\ \emph {et~al.}(1996)\citenamefont {Perdew},
  \citenamefont {Burke},\ and\ \citenamefont {Ernzerhof}}]{pbe96}%
  \BibitemOpen
  \bibfield  {author} {\bibinfo {author} {\bibfnamefont {J.~P.}\ \bibnamefont
  {Perdew}}, \bibinfo {author} {\bibfnamefont {K.}~\bibnamefont {Burke}}, \
  and\ \bibinfo {author} {\bibfnamefont {M.}~\bibnamefont {Ernzerhof}},\ }\href
  {\doibase 10.1103/PhysRevLett.77.3865} {\bibfield  {journal} {\bibinfo
  {journal} {Phys. Rev. Lett.}\ }\textbf {\bibinfo {volume} {77}},\ \bibinfo
  {pages} {3865} (\bibinfo {year} {1996})}\BibitemShut {NoStop}%
\bibitem [{\citenamefont {Bl\"ochl}(1994)}]{blochl1994}%
  \BibitemOpen
  \bibfield  {author} {\bibinfo {author} {\bibfnamefont {P.~E.}\ \bibnamefont
  {Bl\"ochl}},\ }\href {\doibase 10.1103/PhysRevB.50.17953} {\bibfield
  {journal} {\bibinfo  {journal} {Phys. Rev. B}\ }\textbf {\bibinfo {volume}
  {50}},\ \bibinfo {pages} {17953} (\bibinfo {year} {1994})}\BibitemShut
  {NoStop}%
\bibitem [{\citenamefont {Kresse}\ and\ \citenamefont
  {Joubert}(1999)}]{paw_vasp}%
  \BibitemOpen
  \bibfield  {author} {\bibinfo {author} {\bibfnamefont {G.}~\bibnamefont
  {Kresse}}\ and\ \bibinfo {author} {\bibfnamefont {D.}~\bibnamefont
  {Joubert}},\ }\href {\doibase 10.1103/PhysRevB.59.1758} {\bibfield  {journal}
  {\bibinfo  {journal} {Phys. Rev. B}\ }\textbf {\bibinfo {volume} {59}},\
  \bibinfo {pages} {1758} (\bibinfo {year} {1999})}\BibitemShut {NoStop}%
\bibitem [{\citenamefont {Kresse}\ and\ \citenamefont {Hafner}(1993)}]{vasp1}%
  \BibitemOpen
  \bibfield  {author} {\bibinfo {author} {\bibfnamefont {G.}~\bibnamefont
  {Kresse}}\ and\ \bibinfo {author} {\bibfnamefont {J.}~\bibnamefont
  {Hafner}},\ }\href {\doibase 10.1103/PhysRevB.47.558} {\bibfield  {journal}
  {\bibinfo  {journal} {Phys. Rev. B}\ }\textbf {\bibinfo {volume} {47}},\
  \bibinfo {pages} {558} (\bibinfo {year} {1993})}\BibitemShut {NoStop}%
\bibitem [{\citenamefont {Kresse}\ and\ \citenamefont
  {Furthm\"uller}(1996)}]{vasp2}%
  \BibitemOpen
  \bibfield  {author} {\bibinfo {author} {\bibfnamefont {G.}~\bibnamefont
  {Kresse}}\ and\ \bibinfo {author} {\bibfnamefont {J.}~\bibnamefont
  {Furthm\"uller}},\ }\href {\doibase 10.1103/PhysRevB.54.11169} {\bibfield
  {journal} {\bibinfo  {journal} {Phys. Rev. B}\ }\textbf {\bibinfo {volume}
  {54}},\ \bibinfo {pages} {11169} (\bibinfo {year} {1996})}\BibitemShut
  {NoStop}%
\bibitem [{\citenamefont {Madsen}\ \emph {et~al.}(2018)\citenamefont {Madsen},
  \citenamefont {Carrete},\ and\ \citenamefont {Verstraete}}]{madsen2018}%
  \BibitemOpen
  \bibfield  {author} {\bibinfo {author} {\bibfnamefont {G.~K.}\ \bibnamefont
  {Madsen}}, \bibinfo {author} {\bibfnamefont {J.}~\bibnamefont {Carrete}}, \
  and\ \bibinfo {author} {\bibfnamefont {M.~J.}\ \bibnamefont {Verstraete}},\
  }\href {\doibase https://doi.org/10.1016/j.cpc.2018.05.010} {\bibfield
  {journal} {\bibinfo  {journal} {CompAppendixut. Phys. Commun.}\ }\textbf
  {\bibinfo {volume} {231}},\ \bibinfo {pages} {140} (\bibinfo {year}
  {2018})}\BibitemShut {NoStop}%
\bibitem [{\citenamefont {Glazer}(1972)}]{glazer1972}%
  \BibitemOpen
  \bibfield  {author} {\bibinfo {author} {\bibfnamefont {A.~M.}\ \bibnamefont
  {Glazer}},\ }\href {\doibase 10.1107/S0567740872007976} {\bibfield  {journal}
  {\bibinfo  {journal} {Acta Cryst. B}\ }\textbf {\bibinfo {volume} {28}},\
  \bibinfo {pages} {3384} (\bibinfo {year} {1972})}\BibitemShut {NoStop}%
\bibitem [{\citenamefont {Amadon}\ \emph {et~al.}(2008)\citenamefont {Amadon},
  \citenamefont {Lechermann}, \citenamefont {Georges}, \citenamefont {Jollet},
  \citenamefont {Wehling},\ and\ \citenamefont {Lichtenstein}}]{amadon2008}%
  \BibitemOpen
  \bibfield  {author} {\bibinfo {author} {\bibfnamefont {B.}~\bibnamefont
  {Amadon}}, \bibinfo {author} {\bibfnamefont {F.}~\bibnamefont {Lechermann}},
  \bibinfo {author} {\bibfnamefont {A.}~\bibnamefont {Georges}}, \bibinfo
  {author} {\bibfnamefont {F.}~\bibnamefont {Jollet}}, \bibinfo {author}
  {\bibfnamefont {T.~O.}\ \bibnamefont {Wehling}}, \ and\ \bibinfo {author}
  {\bibfnamefont {A.~I.}\ \bibnamefont {Lichtenstein}},\ }\href {\doibase
  10.1103/PhysRevB.77.205112} {\bibfield  {journal} {\bibinfo  {journal} {Phys.
  Rev. B}\ }\textbf {\bibinfo {volume} {77}},\ \bibinfo {pages} {205112}
  (\bibinfo {year} {2008})}\BibitemShut {NoStop}%
\bibitem [{\citenamefont {Sch{\"{u}}ler}\ \emph {et~al.}(2018)\citenamefont
  {Sch{\"{u}}ler}, \citenamefont {Peil}, \citenamefont {Kraberger},
  \citenamefont {Pordzik}, \citenamefont {Marsman}, \citenamefont {Kresse},
  \citenamefont {Wehling},\ and\ \citenamefont {Aichhorn}}]{schuler2018}%
  \BibitemOpen
  \bibfield  {author} {\bibinfo {author} {\bibfnamefont {M.}~\bibnamefont
  {Sch{\"{u}}ler}}, \bibinfo {author} {\bibfnamefont {O.~E.}\ \bibnamefont
  {Peil}}, \bibinfo {author} {\bibfnamefont {G.~J.}\ \bibnamefont {Kraberger}},
  \bibinfo {author} {\bibfnamefont {R.}~\bibnamefont {Pordzik}}, \bibinfo
  {author} {\bibfnamefont {M.}~\bibnamefont {Marsman}}, \bibinfo {author}
  {\bibfnamefont {G.}~\bibnamefont {Kresse}}, \bibinfo {author} {\bibfnamefont
  {T.~O.}\ \bibnamefont {Wehling}}, \ and\ \bibinfo {author} {\bibfnamefont
  {M.}~\bibnamefont {Aichhorn}},\ }\href {\doibase 10.1088/1361-648X/aae80a}
  {\bibfield  {journal} {\bibinfo  {journal} {J. Phys. Condens. Matter}\
  }\textbf {\bibinfo {volume} {30}},\ \bibinfo {pages} {475901} (\bibinfo
  {year} {2018})}\BibitemShut {NoStop}%
\bibitem [{\citenamefont {Blaha}\ \emph {et~al.}(2018)\citenamefont {Blaha},
  \citenamefont {Schwarz}, \citenamefont {Madsen}, \citenamefont {Kvasnicka},
  \citenamefont {Luitz}, \citenamefont {Laskowski}, \citenamefont {Tran},\ and\
  \citenamefont {Marks}}]{Blaha2018}%
  \BibitemOpen
  \bibfield  {author} {\bibinfo {author} {\bibfnamefont {P.}~\bibnamefont
  {Blaha}}, \bibinfo {author} {\bibfnamefont {K.}~\bibnamefont {Schwarz}},
  \bibinfo {author} {\bibfnamefont {G.~K.~H.}\ \bibnamefont {Madsen}}, \bibinfo
  {author} {\bibfnamefont {D.}~\bibnamefont {Kvasnicka}}, \bibinfo {author}
  {\bibfnamefont {J.}~\bibnamefont {Luitz}}, \bibinfo {author} {\bibfnamefont
  {R.}~\bibnamefont {Laskowski}}, \bibinfo {author} {\bibfnamefont
  {F.}~\bibnamefont {Tran}}, \ and\ \bibinfo {author} {\bibfnamefont {L.~D.}\
  \bibnamefont {Marks}},\ }\href
  {http://www.wien2k.at/reg_user/textbooks/usersguide.pdf} {\emph {\bibinfo
  {title} {{WIEN2k, An Augmented Plane Wave + Local Orbitals Program for
  Calculating Crystal Properties}}}}\ (\bibinfo  {publisher} {{K. Schwarz,
  Techn. Univ. Wien, Austria}},\ \bibinfo {year} {2018})\BibitemShut {NoStop}%
\bibitem [{\citenamefont {Aichhorn}\ \emph {et~al.}(2016)\citenamefont
  {Aichhorn}, \citenamefont {Pourovskii}, \citenamefont {Seth}, \citenamefont
  {Vildosola}, \citenamefont {Zingl}, \citenamefont {Peil}, \citenamefont
  {Deng}, \citenamefont {Mravlje}, \citenamefont {Kraberger}, \citenamefont
  {Martins}, \citenamefont {Ferrero},\ and\ \citenamefont
  {Parcollet}}]{TRIQS/DFTTOOLS}%
  \BibitemOpen
  \bibfield  {author} {\bibinfo {author} {\bibfnamefont {M.}~\bibnamefont
  {Aichhorn}}, \bibinfo {author} {\bibfnamefont {L.}~\bibnamefont
  {Pourovskii}}, \bibinfo {author} {\bibfnamefont {P.}~\bibnamefont {Seth}},
  \bibinfo {author} {\bibfnamefont {V.}~\bibnamefont {Vildosola}}, \bibinfo
  {author} {\bibfnamefont {M.}~\bibnamefont {Zingl}}, \bibinfo {author}
  {\bibfnamefont {O.~E.}\ \bibnamefont {Peil}}, \bibinfo {author}
  {\bibfnamefont {X.}~\bibnamefont {Deng}}, \bibinfo {author} {\bibfnamefont
  {J.}~\bibnamefont {Mravlje}}, \bibinfo {author} {\bibfnamefont {G.~J.}\
  \bibnamefont {Kraberger}}, \bibinfo {author} {\bibfnamefont {C.}~\bibnamefont
  {Martins}}, \bibinfo {author} {\bibfnamefont {M.}~\bibnamefont {Ferrero}}, \
  and\ \bibinfo {author} {\bibfnamefont {O.}~\bibnamefont {Parcollet}},\ }\href
  {\doibase 10.1016/j.cpc.2016.03.014} {\bibfield  {journal} {\bibinfo
  {journal} {Comput. Phys. Commun.}\ }\textbf {\bibinfo {volume} {204}},\
  \bibinfo {pages} {200} (\bibinfo {year} {2016})}\BibitemShut {NoStop}%
\bibitem [{\citenamefont {Parcollet}\ \emph {et~al.}(2015)\citenamefont
  {Parcollet}, \citenamefont {Ferrero}, \citenamefont {Ayral}, \citenamefont
  {Hafermann}, \citenamefont {Krivenko}, \citenamefont {Messio},\ and\
  \citenamefont {Seth}}]{TRIQS}%
  \BibitemOpen
  \bibfield  {author} {\bibinfo {author} {\bibfnamefont {O.}~\bibnamefont
  {Parcollet}}, \bibinfo {author} {\bibfnamefont {M.}~\bibnamefont {Ferrero}},
  \bibinfo {author} {\bibfnamefont {T.}~\bibnamefont {Ayral}}, \bibinfo
  {author} {\bibfnamefont {H.}~\bibnamefont {Hafermann}}, \bibinfo {author}
  {\bibfnamefont {I.}~\bibnamefont {Krivenko}}, \bibinfo {author}
  {\bibfnamefont {L.}~\bibnamefont {Messio}}, \ and\ \bibinfo {author}
  {\bibfnamefont {P.}~\bibnamefont {Seth}},\ }\href {\doibase
  http://dx.doi.org/10.1016/j.cpc.2015.04.023} {\bibfield  {journal} {\bibinfo
  {journal} {Comput. Phys. Commun.}\ }\textbf {\bibinfo {volume} {196}},\
  \bibinfo {pages} {398} (\bibinfo {year} {2015})}\BibitemShut {NoStop}%
\bibitem [{yar(2020)}]{yareta}%
  \BibitemOpen
  \href@noop {} {\bibfield  {journal} {\bibinfo  {journal}
  {\href{https://doi.org/10.26037/yareta:jif2uzxdfrgctma6b4rknipqoe}{DOI:10.26037/yareta:jif2uzxdfrgctma6b4rknipqoe}}\
  } (\bibinfo {year} {2020})}\BibitemShut {NoStop}%
\bibitem [{\citenamefont {Fowlie}\ \emph {et~al.}(2017)\citenamefont {Fowlie},
  \citenamefont {Gibert}, \citenamefont {Tieri}, \citenamefont {Gloter},
  \citenamefont {Iniguez}, \citenamefont {Filippetti}, \citenamefont
  {Catalano}, \citenamefont {Gariglio}, \citenamefont {Schober}, \citenamefont
  {Guennou}, \citenamefont {Kreisel}, \citenamefont {Stephan},\ and\
  \citenamefont {Triscone}}]{fowlie2017}%
  \BibitemOpen
  \bibfield  {author} {\bibinfo {author} {\bibfnamefont {J.}~\bibnamefont
  {Fowlie}}, \bibinfo {author} {\bibfnamefont {M.}~\bibnamefont {Gibert}},
  \bibinfo {author} {\bibfnamefont {G.}~\bibnamefont {Tieri}}, \bibinfo
  {author} {\bibfnamefont {A.}~\bibnamefont {Gloter}}, \bibinfo {author}
  {\bibfnamefont {J.}~\bibnamefont {Iniguez}}, \bibinfo {author} {\bibfnamefont
  {A.}~\bibnamefont {Filippetti}}, \bibinfo {author} {\bibfnamefont
  {S.}~\bibnamefont {Catalano}}, \bibinfo {author} {\bibfnamefont
  {S.}~\bibnamefont {Gariglio}}, \bibinfo {author} {\bibfnamefont
  {A.}~\bibnamefont {Schober}}, \bibinfo {author} {\bibfnamefont
  {M.}~\bibnamefont {Guennou}}, \bibinfo {author} {\bibfnamefont
  {J.}~\bibnamefont {Kreisel}}, \bibinfo {author} {\bibfnamefont
  {O.}~\bibnamefont {Stephan}}, \ and\ \bibinfo {author} {\bibfnamefont
  {J.-M.}\ \bibnamefont {Triscone}},\ }\href {\doibase 10.1002/adma.201605197}
  {\bibfield  {journal} {\bibinfo  {journal} {Advanced Materials}\ }\textbf
  {\bibinfo {volume} {29}},\ \bibinfo {pages} {1605197} (\bibinfo {year}
  {2017})}\BibitemShut {NoStop}%
\end{thebibliography}
%
%
%
\end{document}